\documentclass[aps,prd,twocolumn,preprintnumbers,superscriptaddress,nofootinbib,floatfix]{revtex4-1}

\usepackage{amsmath,amssymb}
\usepackage{graphicx}
\usepackage{braket}
\usepackage{units}
\usepackage[hyperfootnotes=false]{hyperref}

\def \vec#1{{\boldsymbol{#1}}}
\newcommand{\hc}{\ensuremath{\text{h.c.}}}

\begin{document}

\preprint{UCI-TR-2019-24}

\title{Inclusive Nucleon Decay Searches as a Frontier of Baryon Number Violation}

\author{Julian Heeck}
\email[]{Julian.Heeck@uci.edu}
\affiliation{Department of Physics and Astronomy, University of California, Irvine\\
Irvine, California, 92697-4575, USA}

\author{Volodymyr Takhistov}
\email[]{vtakhist@physics.ucla.edu}
\affiliation{Department of Physics and Astronomy, University of California, Los Angeles\\
Los Angeles, California, 90095-1547, USA}

\hypersetup{
pdftitle={Inclusive Nucleon Decay Searches as a Frontier of Baryon Number Violation},   
pdfauthor={Julian Heeck, Volodymyr Takhistov}
}

\allowdisplaybreaks

\begin{abstract}

Proton decay, and the decay of nucleons in general, constitutes one of the most sensitive probes of high-scale physics beyond the Standard Model. Most of the existing nucleon decay searches have focused primarily on two-body decay channels, motivated by Grand Unified Theories and supersymmetry. However, many higher-dimensional operators violating baryon number by one unit, $\Delta B = 1$, induce multi-body nucleon decay channels, which have been only weakly constrained thus far. While direct searches for all such possible channels are desirable, they are highly impractical. In light of this, we argue that inclusive nucleon decay searches,  $N \rightarrow X +$anything (where $X$ is a light Standard Model particle with an unknown energy distribution), are particularly valuable, as are model-independent and invisible nucleon decay searches such as $n\to$~invisible. We comment on complementarity and opportunities for such searches in the current as well as upcoming large-scale experiments Super-Kamiokande, Hyper-Kamiokande, JUNO, and DUNE. Similar arguments apply to $\Delta B >1$ processes, which kinematically allow for even more involved final states and are essentially unexplored experimentally.

\end{abstract}

\maketitle

\tableofcontents


\section{Introduction}

Baryon number $B$ and lepton number $L$ are seemingly accidentally conserved in the Standard Model (SM), which makes searches for their violation extremely important. So far we have not observed any $B$ or $L$ violating processes despite decades of experimental investigation, yet there are many reasons to expect that these symmetries could be broken.
The linear combination $B+L$ is in principle already violated by $3+3$ units within the SM itself through non-perturbative instanton effects, albeit highly suppressed~\cite{tHooft:1976rip}.
From a more fundamental top-down perspective, global symmetries such as $U(1)_B$ and $U(1)_L$ are expected to be violated at some level by quantum gravity effects~\cite{Banks:2010zn,Harlow:2018jwu}, which is however difficult to quantify.
Furthermore, baryon number violation is one of the key prerequisites for successful baryogenesis~\cite{Sakharov:1967dj}, which would address the observed baryon--antibaryon asymmetry of our Universe. Explicit $B$-violation and associated proton decay is a defining prediction of Grand Unified Theories (GUTs)~\cite{Georgi:1974sy,Fritzsch:1974nn} that unify the three forces of the SM into a single gauge group, offering an explanation for the observed charge quantization  as well as gauge coupling unification. 
GUTs typically lead to effective dimension-six ($d=6$) operators with $\Delta B = \Delta L =1$ that induce two-body nucleon decays such as $p\to e^+\pi^0$ and $n\to e^+\pi^-$, mediated by heavy gauge bosons with family-universal couplings~\cite{Langacker:1980js,Nath:2006ut,Babu:2013jba}.

It has to be stressed, however, that the significance of nucleon decays stretches far beyond GUTs. $\Delta B\neq 0$ processes generically appear in numerous theoretical extensions of the SM, such as supersymmetry (SUSY)~\cite{Nath:2006ut}, typically mediated at the renormalizable level by leptoquarks and diquarks~\cite{Bowes:1996xy,Kovalenko:2002eh,Arnold:2012sd,Assad:2017iib}.
A somewhat model-independent approach to study $\Delta B \neq 0$ operators is through the SM effective field theory (SMEFT), neglecting model-dependent interference effects~\cite{Weinberg:1979sa,Wilczek:1979hc,Weinberg:1980bf,Weldon:1980gi,Abbott:1980zj}. This allows one to identify higher-dimensional operators with a flavor structure that can make rather unconventional nucleon decay channels dominant.
Further, higher-dimensional $d>6$ operators typically induce nucleon decays with multi-body final states and
even in simple UV-complete models one can encounter more complicated nucleon decay channels such as $n\to e^+ e^-\nu$~\cite{Pati:1983jk} or $p\to \pi^+\pi^+ e^- \nu\nu$~\cite{Weinberg:1980bf,Arnold:2012sd}.
Hence, there is a vast landscape of possible motivated nucleon decay modes of varying complexity.

Experimentally, an extensive nucleon decay search program has been carried out over multiple decades, covering more than $60$ decay channels~\cite{Tanabashi:2018oca}. The most sensitive searches, coming from the Super-Kamiokande (SK) experiment~\cite{Abe:2013gga,Fukuda:2002uc} (see Ref.~\cite{Takhistov:2016eqm} for a review), have already pushed the nucleon lifetime limits for certain channels above $\unit[10^{34}]{yr}$~\cite{Miura:2016krn}, twenty-four orders of magnitude beyond the age of our Universe.
The frontier of baryon number violation searches will be spearheaded by next-generation large-scale underground neutrino experiments, namely the Jiangmen Underground Neutrino Observatory (JUNO)~\cite{Djurcic:2015vqa}, the Deep Underground Neutrino Experiment (DUNE)~\cite{Acciarri:2015uup}, and Hyper-Kamiokande (HK)~\cite{Abe:2018uyc}. It is of paramount importance to take full advantage of these considerable efforts and to identify potential new signals in order to ensure that interesting channels are not overlooked due to theoretical biases.

In this work we revisit nucleon decay channels arising from higher-dimensional operators and discuss some of the possible resulting final states. While systematically searching through all of the kinematically allowed nucleon decay channels with increased final-state complexity would constitute the strongest probes, this approach quickly becomes highly impractical beyond the simplest of the modes. In view of this, we highlight the importance of \emph{inclusive nucleon decay searches}. Although these searches are not as sensitive as exclusive ones looking at a particular channel, they allow one to cover very broad parameter space in a model-independent manner and are practically far more feasible.~This approach is particularly fruitful to revisit in view of the upcoming large-scale experiments.

This paper is organized as follows: 
in Sec.~\ref{sec:theory} we discuss higher-dimensional operators that lead to nucleon decay and argue in particular that many of them lead to multibody final states that are not covered in current searches.
In Sec.~\ref{sec:exclusive_searches} we provide a brief overview of current and upcoming detectors as well as existing exclusive nucleon-decay searches. 
We discuss and propose possible inclusive searches as well as model-independent signatures in Sec.~\ref{sec:inclusive_searches}.
Sec.~\ref{sec:higher_B} is devoted to a short discussion of $\Delta B >1$ processes such as dinucleon decay, which would also profit from inclusive searches.
Finally, we conclude in Sec.~\ref{sec:conclusions}.

\section{Nucleon decay operators}
\label{sec:theory}

Several well-motivated theoretical models such as GUTs or $R$-parity-violating SUSY lead to nucleon decay, typically with specific two-body channels such as $p\to e^+ \pi^0$ or $p\to \bar\nu K^+$ being dominant~\cite{Nath:2006ut,Babu:2013jba}. In order to discuss nucleon decay in its generality without being restricted to a certain model, we instead consider various possible higher-dimensional operators that can mediate these processes, an approach\footnote{This approach does not cover the case of beyond-the-SM light particles $X$ that could lead for example to $p\to \ell^+ + X$~\cite{Takhistov:2015fao}.} that goes back to Refs.~\cite{Weinberg:1979sa,Wilczek:1979hc,Weinberg:1980bf,Weldon:1980gi,Abbott:1980zj}.
We aim to determine which operators lead to two-body and which lead to multi-body nucleon decays.

\subsection{Operator dimension \texorpdfstring{$d=6$}{d=6}}

In the SMEFT, operators that exhibit $\Delta B=1$ start to appear at operator mass dimension $d=6$. Keeping the flavor structure, these $\Delta B=\Delta L =1$ operators can be written as
\begin{align}
\mathcal{L}_{d=6} &= 
y_{abcd}^1 \epsilon^{\alpha\beta\gamma}(\overline{d}^C_{a,\alpha} u_{b,\beta})(\overline{Q}^C_{i,c,\gamma}\epsilon_{ij} L_{j,d}) \nonumber\\
&\quad +y_{abcd}^2 \epsilon^{\alpha\beta\gamma} (\overline{Q}^C_{i,a,\alpha}\epsilon_{ij} Q_{j,b,\beta})(\overline{u}^C_{c,\gamma} \ell_{d}) \nonumber\\
&\quad +y_{abcd}^3 \epsilon^{\alpha\beta\gamma} \epsilon_{il}\epsilon_{jk} (\overline{Q}^C_{i,a,\alpha} Q_{j,b,\beta})(\overline{Q}^C_{k,c,\gamma}  L_{l,d}) \nonumber\\
&\quad +y_{abcd}^4 \epsilon^{\alpha\beta\gamma} (\overline{d}^C_{a,\alpha} u_{b,\beta})(\overline{u}^C_{c,\gamma} \ell_{d})+\text{h.c.} \,,
\label{eq:dequal6}
\end{align}
where  $\alpha,\beta,\gamma$ denote the color, $i,j,k,l$ the $\mathrm{SU}(2)_L$, and $a,b,c,d$ the family indices~\cite{Weinberg:1979sa,Wilczek:1979hc,Weinberg:1980bf,Weldon:1980gi,Abbott:1980zj}.
$u$, $d$, and $\ell$ are the right-handed up-quark, down-quark, and lepton fields, while $Q$ and $L$ are the left-handed quark and lepton doublets, respectively.
The $y^j$ couplings have mass dimension $-2$ and the first-generation entries are constrained to be $<(\unit[\mathcal{O}(10^{15})]{GeV})^{-2}$ due to the induced two-body nucleon decays. Specifically, all of the above operators generate the well-constrained decay $p\to e^+ \pi^0$ with a rate of order
\begin{align}
\Gamma (p\to e^+ \pi^0) \simeq \frac{1}{\unit[2\times 10^{34}]{yr}} \left| \frac{y_{1111}^j}{(\unit[3\times 10^{15}]{GeV})^{-2}}\right|^2 .
\end{align} 
A variety of other two-body nucleon decay channels are induced as well, including muon and kaon modes once we consider second-generation flavor indices. Three-body decay modes with similar rates are induced as well~\cite{Wise:1980ch}  but ultimately lead to weaker constraints.

Operators in $\mathcal{L}_{d=6}$ involving either charm, top, bottom or tau are seemingly unconstrained by nucleon decay since these particles are heavier than the proton; it is however possible to go through heavy \emph{off-shell} particles and still induce nucleon decay, as emphasized in Ref.~\cite{Marciano:1994bg} for operators involving a tau and more generally in Ref.~\cite{Hou:2005iu} (see also Ref.~\cite{Dorsner:2012nq} for a UV-complete example with scalar leptoquarks). As an extreme example, Ref.~\cite{Hou:2005iu} has considered the coupling $y_{3333}^4$ and shown that at two-loop level the simple decay $n\to \bar{\nu}_\tau \pi^0$ is induced with an estimated rate
\begin{align}
\Gamma (n\to \bar{\nu}_\tau \pi^0) \simeq \frac{1}{\unit[10^{33}]{yr}} \left| \frac{y_{3333}^4}{(\unit[5\times 10^{8}]{GeV})^{-2}}\right|^2 .
\end{align} 
An even stronger limit $y_{3333}^4\lesssim (\unit[10^{11}]{GeV})^{-2}$ has been estimated from $p\to \bar{\nu}_\tau K^+$ in Ref.~\cite{Baldes:2011mh}. 
Despite the suppression by loop factors, Cabibbo--Kobayashi--Maskawa mixing angles, and the Fermi constant $G_F$, these limits are far stronger than any constraints from $\Delta B=1$ top or tau decays on the same operator, making nucleon decays clearly the best search channels.
Notice that any operator involving tau leptons brings a missing-energy tau neutrino in the final state, reducing the detection efficiency somewhat; turning this around, it is then crucial to search for (two-body) nucleon decays involving one neutrino, as these are the best channels of $\Delta B=1$ operators that involve tau leptons. As it is conceivable that the UV completion that generates the $\Delta B=1$ operators also singles out tau leptons (or any other lepton flavor for that matter), one must not rely exclusively on searches involving electrons and muons~\cite{Hambye:2017qix}.

The main conclusion of the above exercise is that \emph{any} $\Delta B=1$ operator leads to nucleon decay, no matter the flavor structure. In the $d=6$ case of Eq.~\eqref{eq:dequal6} it is furthermore always possible to close SM loops in order to generate a \emph{two-body} nucleon decay. These might not always be the dominant decay modes, but in light of the clean decay channels they are clearly the preferred way to search for the operators of Eq.~\eqref{eq:dequal6}. This picture changes once we consider $\Delta B=1$ operators of mass dimension $d >6$ as we discuss below.

\subsection{Operator dimension \texorpdfstring{$d>6$}{d>6}}

Assuming a $\Delta B=1$ operator of mass dimension $d\geq 6$ with coefficient $\Lambda^{4-d}$ we can estimate the amplitude for an $k$-body nucleon decay as $\mathcal{M}\sim m_p^{d-k-1} \Lambda^{4-d}$ and the decay width~\cite{Kleiss:1985gy}
\begin{align}
\Gamma (N\to k\text{ particles})\sim \frac{2^{5-4 k} \pi^{3-2 k}}{2 m_p}  \frac{m_p^{2 k-4} |\mathcal{M}|^2}{(k-1)! (k-2)!} \,,
\end{align}
neglecting the final-state masses and possible symmetry factors (see Ref.~\cite{Helo:2019yqp} for loop-suppressed nucleon decays). The decay is suppressed for large $d$  and large $k$, effectively lowering the probed scales $\Lambda$\footnote{We note that the ratio $\Gamma(N \to k_1~\text{particles})/\Gamma (N \to k_2~\text{particles})$ for $k_1 > k_2$ could also be larger than 1 and not follow the suppression just from phase-factor considerations, as shown in Ref.~\cite{Wise:1980ch}.}. Assuming conservatively that $\Lambda$ should lie above TeV in order to evade LHC constraints on the underlying colored mediator particles with quark couplings, we find that $d>14$ (for $k=2$) or $d>6$ (for $k=15$) in order to push the lifetime above $\unit[10^{30}]{yr}$ (which corresponds to a reasonable lower bound on the \emph{total} nucleon lifetime, see discussion in Sec.~\ref{sec:inclusive_searches}).
Clearly, nucleon decays can probe very high-dimensional operators and very high multiplicity, making it possible and necessary to go beyond two-body decays mediated by $d=6$ operators.

\begin{figure*}[t]
	\includegraphics[width=0.8\textwidth]{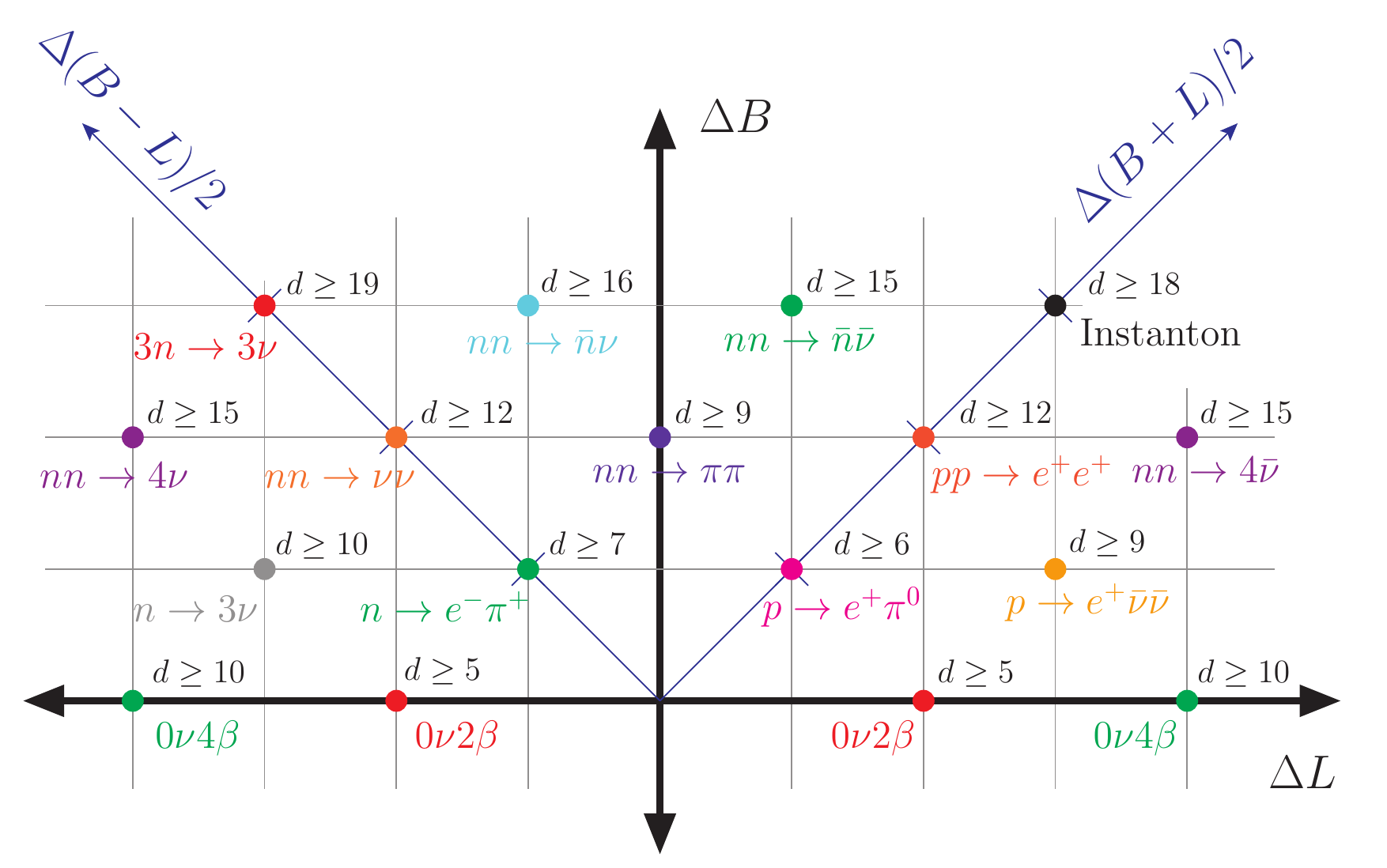}
	\caption{Processes with baryon and lepton number violation by $\Delta B$ and $\Delta L$ units, respectively. We only show one example process, others are implied (e.g.~$nn\to\pi\pi$ also give $n$--$\bar n$ oscillation, $pp\to \pi^+\pi^+$, and many more). 
``Instanton'' refers to processes such as $3n\to 3\bar\nu$ that break the same quantum numbers as non-perturbative electroweak instantons.	$0\nu 2\beta$ ($0\nu 4\beta$) refers to neutrinoless double~\cite{Furry:1939qr,Rodejohann:2011mu} (quadruple~\cite{Heeck:2013rpa}) beta decay.
Final states with neutrinos make an experimental determination of $\Delta L$ impossible, but are shown here for the sake of brevity.
Also shown is the minimal mass dimension $d$ of the underlying effective operator following Ref.~\cite{Helset:2019eyc}.
In addition to total lepton number $L$, all operators and processes also carry lepton \emph{flavor}, which turns the above two-dimensional plot into a four-dimensional lattice~\cite{Heeck:2016xwg,Hambye:2017qix}.
	}
	\label{fig:BL_grid}
\end{figure*}

As already pointed out by Weinberg~\cite{Weinberg:1980bf}, $\Delta B=1$ operators with $d > 6$ can carry different total \emph{lepton} number $\Delta L$, which can be used to make them dominant over the $d=6$ operators. Interesting connections between $\Delta B$, $\Delta L$, and the mass dimension of the operator $d$ were proven in Refs.~\cite{Kobach:2016ami,Helset:2019eyc},
\begin{align}
d \text{ is even} \quad &\leftrightarrow \quad |\Delta (B-L)| = 0, 4, 8, 12, \dots ,\\
d \text{ is odd} \quad &\leftrightarrow \quad |\Delta (B-L)| = 2, 6, 10, 14, \dots  ,
\end{align}
as well as the weak inequality
\begin{align}
d_\text{min}\geq \frac{9}{2}|\Delta B| + \frac{3}{2} |\Delta L| 
\end{align}
for the minimum dimension $d_\text{min}$ of an operator with $\Delta B$ and $\Delta L$. For the cases of practical interest we give $d_\text{min}$ in Fig.~\ref{fig:BL_grid}, adapted from Ref.~\cite{Helset:2019eyc}.
Below we discuss $\Delta B=1$ operators with $d>6$ in order to establish the importance of multi-body nucleon decay searches.

At $d=7$ one finds $\Delta B=-\Delta L =1$ operators that induce e.g.~$n\to e^-K^+$ or $p\to e^- \pi^+ K^+$~\cite{Weinberg:1980bf}. In analogy to the $d=6$ operators from above one can show that all of these $d=7$ operators induce \emph{two-body} nucleon decays at some loop level, irrespective of their flavor structure. While these might not necessarily be the dominant decay modes, they are clearly far easier to constrain experimentally. The currently best constrained channels are $p\to \nu K^+$ and $n\to \ell^- \pi^+$, while many other two-body final states with $|\Delta (B-L)|=2$ have unfortunately not been updated for twenty years (listed below in Tab.~\ref{tab:two-body_limits}).

At $d=8$ one finds $\Delta B=\Delta L=1$ operators of the form $uuQL \bar\phi\bar\phi$, $ddQL\phi\phi$, and $dQQ\ell \phi\phi$ in addition to simply dressing the $d=6$ operators of Eq.~\eqref{eq:dequal6} with $|\phi|^2$, where $\phi$ is the SM scalar doublet.\footnote{Here and in the following we ignore operators that contain derivatives. A full list of these and other operators can be conveniently obtained using the program \texttt{Sym2Int}~\cite{Fonseca:2017lem,Fonseca:2019yya}.} Once again two-body nucleon decays are the best search channels.

Starting at $d=9$ the phenomenology becomes vastly more interesting. Suppressing all indices and using a very compact notation, the $\Delta B=1$ operators can be written in the form
\begin{align}
\mathcal{O}^9_1 &= d  d  d  \ell \bar{\ell}  \bar{\ell}\,, &
\mathcal{O}^9_2 &= u  d  d  \ell \bar{L}  \bar{L}\,,\nonumber \\
\mathcal{O}^9_3 &= d  d  d  \bar{\ell}  L  \bar{L}\,, &
\mathcal{O}^9_4 &= d  d  Q  \ell \bar{\ell}  \bar{L}\,,\nonumber \\
\mathcal{O}^9_5 &= d  d  Q  L  \bar{L}  \bar{L}\,, &
\mathcal{O}^9_6 &= d  Q  Q  \ell \bar{L}  \bar{L}\,,\nonumber \\
\mathcal{O}^9_7 &= u  d  d  \bar{L}  \bar{\phi}  \bar{\phi}  \phi\,, &
\mathcal{O}^9_8 &= u  Q  Q  \bar{L}  \bar{\phi}  \bar{\phi}  \bar{\phi}\,,\nonumber \\
\mathcal{O}^9_9 &= d  d  d  \bar{L}  \bar{\phi}  \phi  \phi\,, &
\mathcal{O}^9_{10} &= d  d  Q  \bar{\ell}  \bar{\phi}  \bar{\phi}  \phi\,,\nonumber \\
\mathcal{O}^9_{11} &= d  Q  Q  \bar{L}  \bar{\phi}  \bar{\phi}  \phi\,, &
\mathcal{O}^9_{12} &= Q  Q  Q  \bar{\ell}  \bar{\phi}  \bar{\phi}  \bar{\phi}\,, \\
\mathcal{O}^9_{13} &= \bar{u}  u  d  d  d  \bar{\ell}\,, &
\mathcal{O}^9_{14} &= \bar{u}  u  d  d  Q  \bar{L}\,,\nonumber \\
\mathcal{O}^9_{15} &= \bar{u}  d  d  Q  Q  \bar{\ell}\,, &
\mathcal{O}^9_{16} &= \bar{u}  d  Q  Q  Q  \bar{L}\,,\nonumber \\
\mathcal{O}^9_{17} &= u  d  d  d  \bar{Q}  \bar{L}\,, &
\mathcal{O}^9_{18} &= \bar{d}  d  d  d  d  \bar{\ell}\,,\nonumber \\
\mathcal{O}^9_{19} &= \bar{d}  d  d  d  Q  \bar{L}\,, &
\mathcal{O}^9_{20} &= d  d  d  \bar{Q}  Q  \bar{\ell}\,,\nonumber \\
\mathcal{O}^9_{21} &= d  d  \bar{Q}  Q  Q  \bar{L}\,, &
\mathcal{O}^9_{22} &= u  u  u  \ell L  L\,,\nonumber \\
\mathcal{O}^9_{23} &= u  u  Q  L  L  L\,.  &  \nonumber 
\end{align}
$\mathcal{O}^9_1$--$\mathcal{O}^9_6$ have recently been discussed in Ref.~\cite{Hambye:2017qix}; they fulfil $\Delta B= - \Delta L=1$, but since they involve three lepton fields they can carry non-trivial lepton \emph{flavor} number, i.e.~$|\Delta L_\alpha| >1$ for one $\alpha\in \{e,\mu,\tau\}$, which makes it possible to make them dominant over $d=7$ operators (which carry at most $|\Delta L_\alpha| =1$) and then lead to three-body~\cite{ODonnell:1993kdg} and four-body nucleon decays~\cite{Hambye:2017qix}. 
$\mathcal{O}^9_7$--$\mathcal{O}^9_{21}$ leads to $\Delta B=- \Delta L=1$ two-body nucleon decays reminiscent of the $d=7$ operators, although
$\mathcal{O}^9_{13}$--$\mathcal{O}^9_{21}$ dominantly induces multi-meson final states, e.g.~$n\to\ell^- K^+ \pi^0$, due to the large number of quark fields.
Finally, $\mathcal{O}^9_{22}$ and $\mathcal{O}^9_{23}$ violate $\Delta B = \tfrac{1}{3}\Delta L =1$~\cite{Weinberg:1980bf} and thus lead to nucleon decays with at least three antileptons in the final state, typically accompanied by one or more mesons. We stress that contrary to many statements in the literature these operators do indeed generate nucleon decays despite the fact that they necessarily contain charm or top quarks, in complete analogy to the discussion in Ref.~\cite{Hou:2005iu}.
At loop level the operators $\mathcal{O}^9_{22,23}$ with arbitrary flavor structure induce the decays $p\to \ell^+\bar\nu\bar\nu$ and $n\to 3\bar\nu$, which are experimentally constrained already.

At $d=10$ there are $\Delta B=\Delta L =1$ as well as $\Delta B=- \tfrac{1}{3}\Delta L =1$ operators. The former are generically expected to be suppressed compared to $d=6$ operators, except for operators that involve three lepton fields with non-trivial flavor, in particular those operators that give rise to the very clean channels $p\to e^+\mu^-\mu^-$ and $p\to \mu^+ e^- e^-$, discussed at length in Ref.~\cite{Hambye:2017qix}.
The $\Delta B=- \tfrac{1}{3}\Delta L =1$ operators are of the form $ddd\bar L\bar L \bar L \bar \phi$ and lead to multi-body nucleon decays such as $n\to \ell^-\nu\nu K^+$, $p\to \ell^-\nu\nu\pi^+\pi^+$~\cite{Weinberg:1980bf,Arnold:2012sd,Fonseca:2018ehk}, and $n\to 3\nu$, only the latter being explicitly constrained so far.

For examples of $\Delta B =1$ operators with $d>10$ we refer to Refs.~\cite{Appelquist:2001mj,Fonseca:2018ehk,Helset:2019eyc}. If such operators are to dominate over lower-dimensional ones they should carry a different lepton (flavor) number, e.g.~$|\Delta L |>3$. It is clear that these lead to multi-lepton final states, typically accompanied by mesons. 
We note that the semi-inclusive invisible neutron decay, $n\to \text{neutrinos}$, can carry away an arbitrary amount of lepton number and flavor via neutrinos and is thus a particularly powerful decay to probe, not least because the experimental signature is the same no matter the number or flavor of the final-state neutrinos. We come back to this channel in Sec.~\ref{sec:inclusive_searches}.

The calculation and comparison of all possible nucleon final states induced by a given $d\gg 6$ operator is of course impractical and is not attempted here. From the examples discussed above we can however already glean our simple main point: current nucleon decay searches cover but a fraction of relevant modes and should by all means be extended in order not to miss new physics.
From our discussion it is clear that two-body nucleon decays are powerful probes of $d=6$--$8$ operators, whereas $d\geq 9$ operators, in particular those with $|\Delta L|>1$ or $|\Delta L_\alpha|>1$, lead to multi-body final states that mostly lie outside of current searches. Since nucleon decay is an extremely sensitive probe of new physics, even higher-dimensional operators with $d\gg 6$ can give testable signals and should be investigated. To constrain these operators one then has to either go through all kinematically allowed multi-body final states (in general without knowing the angular and spectral distributions) or focus on \emph{inclusive} searches, to be discussed in Sec.~\ref{sec:inclusive_searches}.
Before, we must first discuss the status of \emph{exclusive} searches in Sec.~\ref{sec:exclusive_searches}.

\subsection{UV-complete example}

To illustrate the above discussion we give one simple UV-complete example for a potentially dominant $\Delta B=1$ process that is not covered by existing exclusive searches, closely following  Ref.~\cite{Hambye:2017qix}. We consider the $d = 9$ operator $\mathcal{O}^9_1$ from above, which includes a $(dd)$ pair that is antisymmetric in the flavor indices and thus contains a strange quark. We focus on one particularly interesting lepton flavor combination
\begin{align}
\mathcal{O}^9 = (d  s) (d  \mu) (\bar{e}  \bar{e})/\Lambda^5 \,.
\end{align} 
Of the four globally conserved quantum numbers of the SM, $(B,L_e,L_\mu,L_\tau)$, the operator $\mathcal{O}^9$ breaks the linear combination  $\Delta (B -2 L_e +L_\mu ) =6$ but still conserves $B+L_e + L_\mu$, $L_e + 2 L_\mu$, and $L_\tau$. In fact, it is the lowest-dimensional operator with these properties, which is the reason it can be the dominant nucleon-decay operator.

\begin{figure}[t]
	\includegraphics[width=0.45\textwidth]{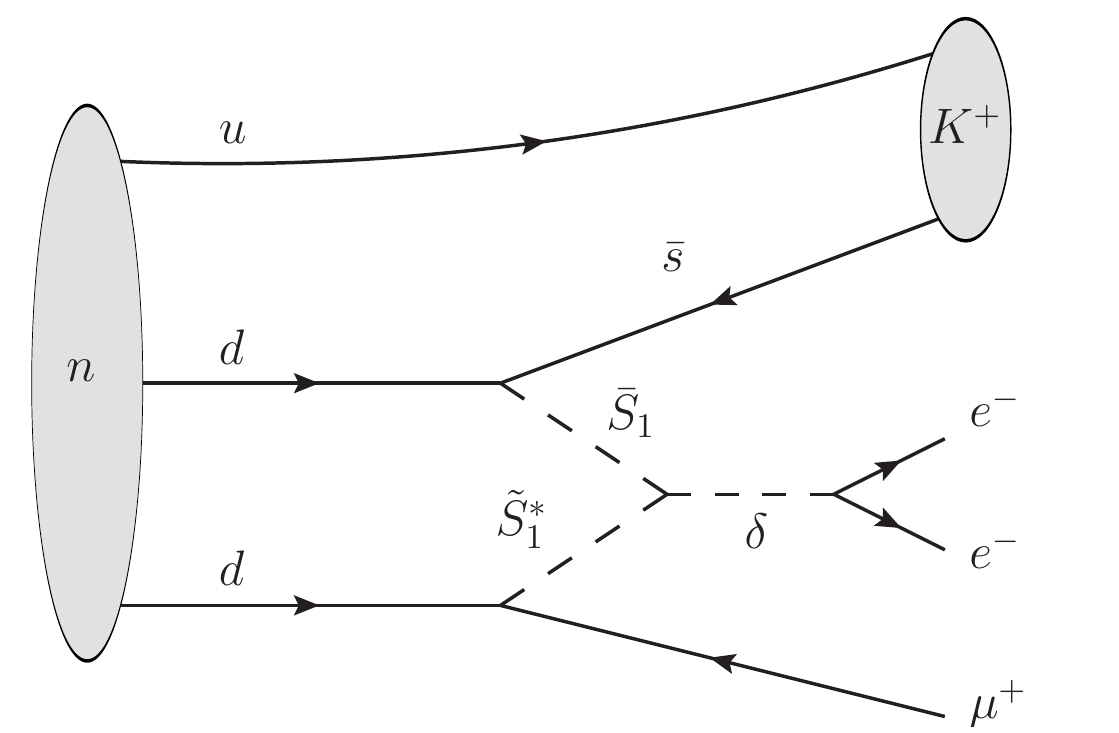}
	\caption{Neutron decay $n\to K^+ \mu^+ e^- e^-$ in the UV-complete example.
	}
	\label{fig:neutron_decay_d9}
\end{figure}

As a UV completion we introduce the two scalar leptoquarks $\bar{S}_1 \sim (\bar{\vec{3}},\vec{1},-2/3)$ and $\tilde{S}_1 \sim (\bar{\vec{3}},\vec{1},4/3)$ as well as the dilepton scalar $\delta \sim (\vec{1},\vec{1},-2)$, with the relevant interactions
\begin{align}
\mathcal{L} =~ &y_1 dd \bar{S}_1^* + y_2 d\ell \tilde{S}_1 + y_3 u u \tilde{S}_1^* + y_4 \ell\ell \delta^*\nonumber\\
&+ \kappa \tilde{S}_1 \bar{S}_1^* \delta + \hc ,
\end{align}
with Yukawa coupling matrices $y_j$ and a scalar-potential coupling constant $\kappa$ with units of mass~\cite{KlapdorKleingrothaus:2002yq}. While we suppress the flavor indices, let us note that $y_1$ and $y_3$ are antisymmetric and $y_4$ symmetric. 
Assigning $L_e(\delta)=2$, $B(\bar{S}_1)=2/3$, and $L_\mu (\tilde{S}_1)= 3 B(\tilde{S}_1)=-1$ and imposing the symmetries conserved by $\mathcal{O}^9$ we enforce $y_3=0$, $y_2 = (y_2)_{j\mu}$, $y_4=(y_4)_{ee}$. This leads to the desired $\mathcal{O}^9$ operator with the effective suppression scale
\begin{align}
\frac{1}{\Lambda^5} \sim \frac{\kappa^* (y_1)_{12} (y_2)_{1\mu} (y_4)_{ee}^*}{m_{\bar{S}_1}^2 m_{\tilde{S}_1}^2 m_{\delta}^2 } 
\end{align}
upon integrating out the new scalars, as illustrated by the Feynman diagram for $n\to K^+ \mu^+ e^- e^-$ in Fig.~\ref{fig:neutron_decay_d9}.
The imposed symmetries ensure that all $\Delta B\neq 0$ processes have to go through $\mathcal{O}^9$, as there is no lower-dimensional operator with the same quantum numbers. While it is possible to attach SM interactions to Fig.~\ref{fig:neutron_decay_d9} to generate three-body final states such as $n\to \mu^+ e^- \nu_e$, these are further suppressed by $G_F$. From neutrino oscillations we already know that the lepton flavor symmetries we imposed are actually broken, which calls into question our usage of them. However, if these symmetries are only broken in the neutrino sector then the flavor-breaking effects in $\Delta B \neq 0$ processes are suppressed by $m_\nu$ and thus negligible~\cite{Hambye:2017qix}.

\begin{figure}[t]
	\includegraphics[width=0.5\textwidth]{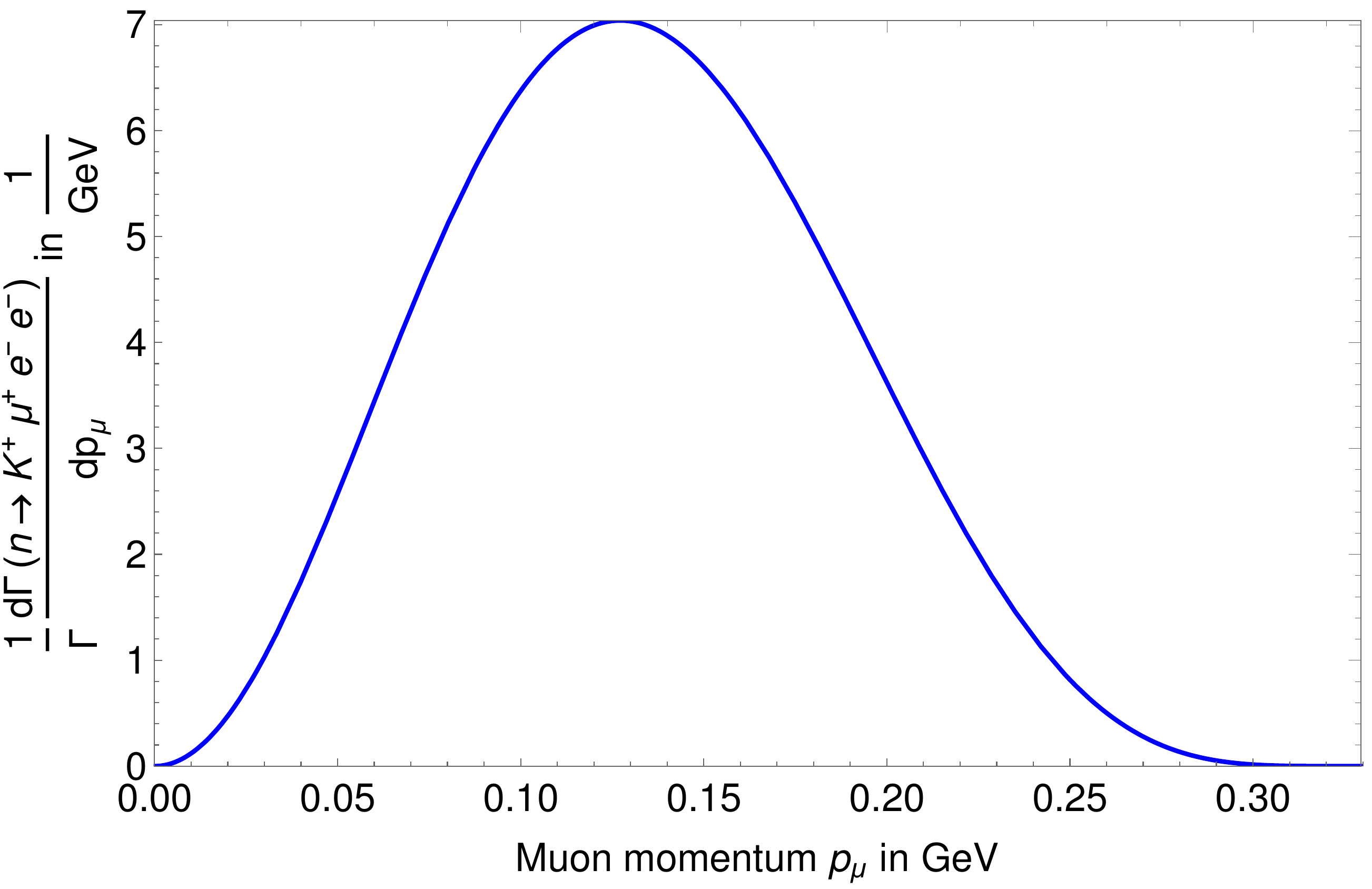}
	\caption{Primary final-state muon spectrum of neutron decay $n\to K^+ \mu^+ e^- e^-$.
	}
	\label{fig:muon_spectrum}
\end{figure}

Having motivated the existence of the operator $\mathcal{O}^9$, we can calculate the total decay rate for $n \to K^+ \mu^+ e^- e^-$. A simple analytic expression can only be obtained if we neglect all final-state masses
\begin{align}
\Gamma (n \to K^+ \mu^+ e^- e^-) \simeq \frac{m_n^7 W_0^2}{737280 \pi^5 \Lambda^{10}}\,,
\end{align}
where $W_0\simeq (\unit[0.23]{GeV})^2$ comes from the relevant nuclear matrix element $\braket{K^+ | dsd | n}$~\cite{Aoki:2017puj}. A numerical evaluation including the kaon and muon masses yields instead
\begin{align}
\Gamma (n \to K^+ \mu^+ e^- e^-) \simeq \frac{1}{\unit[5\times 10^{31}]{yr}}\left(\frac{\unit[100]{TeV}}{\Lambda}\right)^{10} .
\end{align}
The final-state electrons are typically above the Cherenkov threshold, while the kaon is below and thus invisible in experiments such as SK. The muon momentum spectrum is shown in Fig.~\ref{fig:muon_spectrum}, which leads to roughly half of the muons above and half below the Cherenkov threshold.
No dedicated exclusive search for this (or similar) four-body decay has ever been performed, although it has been proposed long ago~\cite{Pati:1983jk}.
Nevertheless, constraints on $\Gamma (n \to K^+ \mu^+ e^- e^-)$ can be obtained from \emph{inclusive} searches, discussed extensively in Sec.~\ref{sec:inclusive_searches}. 
In particular, a limit of
$\tau (N\to\mu^++\text{anything}) > \unit[12\times 10^{30}]{yr}$~\cite{Tanabashi:2018oca} applies, neglecting any detection efficiency penalties. As we argue below, inclusive searches in SK should be able to improve this bound by more than an order of magnitude, with even more progress expected with the future HK. While a dedicated search for $n \to K^+ \mu^+ e^- e^-$ would of course provide the best possible limit, it is clearly not feasible to systematically study all possible multi-body nucleon decay channels. As we show in the next section, even restricting the search to two- and three-body final states is a challenge. \emph{Inclusive} searches on the other hand provide a simpler handle on the final-state complexity and deserve more attention.

\section{Exclusive nucleon decay searches}
\label{sec:exclusive_searches}

Several major directions have been pursued to experimentally study nucleon decays. The main points of focus are the ability of an experiment to 
achieve high signal-detection efficiency as well as scalability to large fiducial masses/volumes, which is particularly critical for such rare-event searches. Some of the earliest searches were performed with scintillators, pioneered by Reines~et.~al.~\cite{Reines:1954pg, Reines:1958pf,Giamati:1962pe,Backenstoss1960,Reines:1976pv}. These exclusive searches, focusing on some specific nucleon decay channels, assumed that a final state of the decay contains an electromagnetically interacting particle. Current large-volume scintillator neutrino experiments based on carbon $^{12}$C, such as KamLAND~\cite{Abe:2008aa} and Borexino~\cite{Alimonti:2008gc}, have fiducial masses of $\lesssim \unit[1]{kton}$ and can be also utilized for nucleon-decay searches (e.g.~Refs.~\cite{Back:2003wj,TheKamLAND-Zen:2015eva}). The main distinguishing feature of these experiments is a very low sub-MeV energy threshold.

Other early searches looked for fragments of fission induced by nucleon decays within radioactive ore. As we further discuss below, a particular advantage of such searches is that they are completely model independent. However, the amount of source material is limited and hence they are significantly less sensitive than dedicated state-of-the-art exclusive searches.
For exclusive searches, fine-grained (typically iron-based) calorimeters, such as in the Kolar Gold Field (KGF)~\cite{Krishnaswamy:1986gp}, NUSEX~\cite{Battistoni:1982vv}, Soudan~\cite{Bartelt:1986cv}, and Fr{\'e}jus~\cite{Berger:1991fa} experiments, have been also utilized. In these experiments one can achieve high signal efficiency due to precise reconstruction and tracking. However, these configurations do not scale well and hence suffer from a limited fiducial mass ($\lesssim \unit[1]{kton}$). 

The most sensitive searches come from large water-Cherenkov (WC) experiments such as Homestake~\cite{Cherry:1981uq}, Harvard--Purdue--Wisconsin (HPW)~\cite{Phillips:1989dp}, Irvine--Michigan--Brookhaven (IMB)~\cite{McGrew:1999nd}, and Kamiokande~\cite{Hirata:1989kn}. WC experiments can identify charged particles with high efficiency and are readily scalable, enabling fiducial volumes far greater than any other technique. The current most sensitive proton-decay limits, already exceeding lifetimes of $\unit[10^{34}]{yr}$, come from the state-of-the-art WC experiment SK~\cite{Miura:2016krn}, which boasts a fiducial volume of \unit[22.5]{kton} and has been collecting data for two decades.

WC detection relies on the presence of electromagnetically interacting particles in the final state of nucleon decays. The observable signature from a particle traversing the detector is a Cherenkov radiation ring, classified either as ``showering'' for $e^{\pm}$ and $\gamma$ or ``non-showering'' for $\mu^{\pm}$ and $\pi^{\pm}$. WC searches are particularly sensitive when the final state is fully visible and one can efficiently reconstruct the original parent nucleon, such as in the case of the leading GUT modes $p \rightarrow e^+ \pi^0$ and $p \rightarrow \mu^+ \pi^0$, where the pion is identified via $\pi^0 \rightarrow \gamma\gamma$.  
To produce Cherenkov radiation, charged particles of mass $m$ traveling in a medium of refraction index $n$ must exceed the Cherenkov momentum threshold of $p_{\rm th} = m/\sqrt{n^2-1}$. In water, $n \simeq 1.33$ and thus $p_{\rm th} = 1.14 m$, which translates to a required minimum momentum of $\unit[0.58]{MeV}$, $\unit[121]{MeV}$, $\unit[159]{MeV}$, and $\unit[563]{MeV}$ for $e^{\pm}$, $\mu^{\pm}$, $\pi^{\pm}$, and $K^{\pm}$, respectively. Hence, kaons from single nucleon decays, which have energy below $m_p/2 \simeq \unit[470]{MeV}$ where $m_p$ is the proton mass, are always invisible in WC detectors and one must rely on reconstructing products from their subsequent decays. Thus the typically leading SUSY GUT decay channel $p \rightarrow \nu K^+$ suffers from a low detection efficiency in WC detectors.
Scintillation detectors have an advantage over WC in that there is no Cherenkov threshold and a higher light yield. However, because emission of scintillation light is nearly isotropic the directional information is lost. Further, while scintillators are scalable, they are still behind in volume compared to leading WC experiments.  

Several next-generation large-scale experiments will allow us to push nucleon decay searches even further. The upcoming successor of SK, the HK WC experiment, is expected to have a $\unit[187]{kton}$ fiducial volume in its initial configuration~\cite{Abe:2018uyc}. This will allow HK to probe nucleon lifetimes up to $\unit[10^{35}]{yr}$. Since a large fiducial volume benefits all the decay modes, improved limits across the board are expected. On the scintillator front, the liquid scintillator experiment JUNO will have \unit[20]{kton} fiducial volume~\cite{Djurcic:2015vqa} and will start taking data in a few years. The  DUNE~\cite{Acciarri:2015uup} experiment with $\sim\unit[40]{kton}$ fiducial volume based on liquid-argon time-projection-chamber (LArTPC) technology will
allow one to track both light deposits as well as charge. Despite their smaller size compared to HK, the JUNO and DUNE detectors are particularly promising for decay modes involving kaons, such as $p \rightarrow \nu K^+$, which are fully visible and can be reconstructed with high efficiency. Additionally, for the multi-body decay channels discussed in this article the efficiency in WC experiments could be decreased, as it becomes difficult to reconstruct multiple overlapping Cherenkov rings, making alternative technologies as those in JUNO and DUNE crucial. For example, in $\Delta B=2$ $n$--$\overline{n}$ oscillations the resulting $\overline{n}$ is subsequently captured on $n$ or $p$ and produces a slew of decaying pions~\cite{Abe:2011ky,Barrow:2019viz}. In DUNE, the multi-track nature of such decays could be utilized with high efficiency~\cite{Hewes:2017xtr}.

We briefly note that planned next-generation dark-matter direct-detection experiments such as Argo~\cite{Aalseth:2017fik} (based on liquid Argon) or Darwin~\cite{Aalbers:2016jon} (Xenon) could be able to achieve ultra-low sub-keV energy thresholds combined with a $\unit[\mathcal{O}(10\text{--}100)]{ton}$ fiducial mass and could in principle also be utilized for nucleon decay searches. However, their fiducial volume is still multiple orders of magnitude smaller than dedicated large-scale neutrino experiments and would have a hard time competing with their exclusive searches, although their low threshold could be beneficial for model-independent searches.

Despite significant experimental efforts proton decay or any other $\Delta B\neq 0$ process has not been observed thus far.
Almost all kinematically allowed two-body nucleon decay channels have been searched for in various experiments, although several of the limits are outdated. We collect the strongest limits in Tab.~\ref{tab:two-body_limits} and also specify the violation in $(B-L)$ symmetry that such decays induce. We note that a neutrino or an antineutrino of any flavor in the final state will escape the detector before interacting and is hence invisible, which can result in a variation of $|\Delta(B-L)|$ by 2 units. We do not include $\Delta B=1$ multi-nucleon decays such as $pn \to e^+ n$ or $pp\to e^+\Delta^+$ here, some of which were searched for in Ref.~\cite{Berger:1991fa}. We also omit showing processes that violate Lorentz or charge symmetry.

\begin{table}[htbp!]
	\begin{center}
		\begin{tabular}{lcr}
			Channel & $|\Delta(B-L)|$ & $\frac{\Gamma^{-1}}{\unit[10^{30}]{yr}}$ \\ \hline\hline
			$p\to e^++\gamma$ & 0 &  $41000$ \cite{Sussman:2018ylo}\\ \hline
			$p\to e^++\pi ^0$ & 0 & $16000$ \cite{Miura:2016krn} \\ \hline
			$p\to e^++\eta$ & 0 & $10000$ \cite{TheSuper-Kamiokande:2017tit}\\ \hline
			$p\to e^++\rho ^0$ & 0 & $720$ \cite{TheSuper-Kamiokande:2017tit} \\ \hline
			$p\to e^++\omega$ & 0 & $1600$ \cite{TheSuper-Kamiokande:2017tit} \\ \hline
			$p\to e^++K^0$ & 0 & $1000$ \cite{Kobayashi:2005pe} \\ \hline
			$p\to e^++K^{*,0}$ & 0 & $84$ \cite{McGrew:1999nd} \\ \hline
			$p\to \mu ^++\gamma$ & 0 & $21000$ \cite{Sussman:2018ylo} \\ \hline
			$p\to \mu ^++\pi ^0$ & 0 &  $7700$ \cite{Miura:2016krn} \\ \hline
			$p\to \mu ^++\eta$ & 0 & $4700$ \cite{TheSuper-Kamiokande:2017tit} \\ \hline
			$p\to \mu ^++\rho ^0$ & 0 & $570$ \cite{TheSuper-Kamiokande:2017tit} \\ \hline
			$p\to \mu ^++\omega$ & 0 & $2800$ \cite{TheSuper-Kamiokande:2017tit} \\ \hline
			$p\to \mu ^++K^0$ & 0 & $1600$ \cite{Regis:2012sn} \\ \hline
			$p\to \nu+\pi ^+$ & 0,2 & $390$ \cite{Abe:2013lua} \\ \hline
			$p\to \nu+\rho ^+$ & 0,2 & $162$ \cite{McGrew:1999nd} \\ \hline
			$p\to \nu+K^+$ & 0,2 & $5900$ \cite{Abe:2014mwa} \\ \hline
			$p\to \nu+K^{*,+}$ & 0,2 & $130$ \cite{Scott:2007} \\ \hline
			\hline
			$n\to e^-+\pi ^+$ & 2 & $65$~\cite{Seidel:1988ut} ($5300^*$~\cite{TheSuper-Kamiokande:2017tit})  \\ \hline
			$n\to e^-+\rho ^+$ & 2 & $62$~\cite{Seidel:1988ut} ($217^*$ \cite{McGrew:1999nd})  \\ \hline
			$n\to e^-+K^+$ & 2 & $32$ \cite{Berger:1991fa} \\ \hline
			$n\to e^-+K^{*,+}$ & 2 &  \\ \hline
			$n\to e^++\pi ^-$ & 0 & $5300$ \cite{TheSuper-Kamiokande:2017tit} \\ \hline
			$n\to e^++\rho ^-$ & 0 & $217$ \cite{McGrew:1999nd} \\ \hline
			$n\to e^++K^-$ & 0 & $17$ \cite{McGrew:1999nd} \\ \hline
			$n\to e^++K^{*,-}$ & 0 &  \\ \hline
			$n\to \mu ^-+\pi ^+$ & 2 & $49$ \cite{Seidel:1988ut} ($3500^*$ \cite{TheSuper-Kamiokande:2017tit}) \\ \hline
			$n\to \mu ^-+\rho ^+$ & 2 & $7$ \cite{Seidel:1988ut} ($228^*$ \cite{McGrew:1999nd}) \\ \hline
			$n\to \mu ^-+K^+$ & 2 & $57$ \cite{Berger:1991fa} \\ \hline
			$n\to \mu ^++\pi ^-$ & 0 & $3500$ \cite{TheSuper-Kamiokande:2017tit} \\ \hline
			$n\to \mu ^++\rho ^-$ & 0 & $228$ \cite{McGrew:1999nd} \\ \hline
			$n\to \mu ^++K^-$ & 0 &  $26$ \cite{McGrew:1999nd} \\ \hline
			$n\to \nu+\gamma$ & 0,2 & $550$ \cite{Takhistov:2015fao} \\ \hline
			$n\to \nu+\pi ^0$ & 0,2 & $1100$ \cite{Abe:2013lua} \\ \hline
			$n\to \nu+\eta$ & 0,2 & $158$ \cite{McGrew:1999nd} \\ \hline
			$n\to \nu+\rho ^0$ & 0,2 & $19$ \cite{Seidel:1988ut} \\ \hline
			$n\to \nu+\omega$ & 0,2 & $108$ \cite{McGrew:1999nd} \\ \hline
			$n\to \nu+K^0$ & 0,2 & $130$ \cite{Kobayashi:2005pe} \\ \hline
			$n\to \nu+K^{*,0}$ & 0,2 & $78$ \cite{McGrew:1999nd} \\ \hline
		\end{tabular}
	\end{center}
	\caption{Kinematically allowed two-body nucleon decays with $90\%$~C.L.~upper limit on the partial decay width $\Gamma$. Here, $\nu$ can be a neutrino or antineutrino of any flavor, which does not change the observation signature. The column $|\Delta(B-L)|$ indicates the violation of $B-L$ in the decay, which depends on whether $\nu$ is a neutrino or an antineutrino.
		An asterisk denotes a limit that has been translated using the properties of WC detectors ($e^-\sim e^+\sim \gamma$, $\mu^-\sim \mu^+$, $\pi^-\sim \pi^+$) but was not given explicitly in the references. 
Unconstrained channels are still subject to limits from \emph{inclusive} searches, discussed in Sec.~\ref{sec:inclusive_searches}.		
		See text for more details.
		\label{tab:two-body_limits}}
\end{table}

In two-body decay modes, kinematics and phase-space considerations uniquely determine the resulting energy-momentum distribution of the resulting final-state particles, making these searches highly model independent. For multi-body decays, additional model dependence from dynamics comes into play, as discussed for example in Ref.~\cite{Chen:2014ifa}. Multi-body searches can furthermore suffer from lowered detection efficiencies and enhanced systematic uncertainties, e.g.~from multiple hadronic/nuclear interactions. Only a subset of all kinematically allowed three-body nucleon decay channels have been searched for so far, collected in Tabs.~\ref{tab:three-body_proton_limits} and~\ref{tab:three-body_neutron_limits} along with the $|\Delta (B-L)|$ structure for the modes. To remain agnostic regarding theoretical models, a uniform phase-space energy and momentum distribution for final-state particles is typically assumed in such searches.

We are not aware of any searches involving more than three particles in the final state so we do not provide a table listing all kinematically allowed modes.
Instead, we discuss \emph{inclusive} searches, which could in principle be sensitive to arbitrary multi-body final states.

\begin{table}[htbp!]
	\begin{center}
		\begin{tabular}{lcr}
			Channel & $|\Delta(B-L)|$ & $\frac{\Gamma^{-1}}{\unit[10^{30}]{yr}}$ \\ \hline\hline
			$p\to e^-+e^+ + e^+$ & 0 & 793 \cite{McGrew:1999nd} \\ \hline
			$p\to e^-+e^++\mu ^+$ & 0 & 529 \cite{McGrew:1999nd} \\ \hline
			$p\to e^+ + e^++\mu ^-$ & 0 & $529^*$ \cite{McGrew:1999nd} \\ \hline
			$p\to e^-+\mu^+ + \mu^+$ & 0 & 6 \cite{Phillips:1989dp} ($359^*$ \cite{McGrew:1999nd}) \\ \hline
			$p\to e^++\mu ^-+\mu ^+$ & 0 & 359 \cite{McGrew:1999nd} \\ \hline
			$p\to \mu ^-+\mu^+ + \mu^+$ & 0 & 675 \cite{McGrew:1999nd} \\ \hline
			$p\to e^++2 \nu$ & 0,2 & 170 \cite{Takhistov:2014pfw} \\ \hline
			$p\to \mu ^++2 \nu$ & 0,2 & 220 \cite{Takhistov:2014pfw} \\ \hline
			$p\to e^-+2 \pi ^+$ & 2 & 30 \cite{Berger:1991fa} ($82^*$ \cite{McGrew:1999nd}) \\ \hline
			$p\to e^-+\pi ^++\rho ^+$ & 2 &   \\ \hline
			$p\to e^-+K^++\pi ^+$ & 2 & 75 \cite{McGrew:1999nd} \\ \hline
			$p\to e^++2 \gamma$ & 0 & 100 \cite{Berger:1990kg} ($793^*$ \cite{McGrew:1999nd}) \\ \hline
			$p\to e^++\pi ^-+\pi ^+$ & 0 & 82 \cite{McGrew:1999nd} \\ \hline
			$p\to e^++\rho ^-+\pi ^+$ & 0 &   \\ \hline
			$p\to e^++K^-+\pi ^+$ & 0 & $75^*$ \cite{McGrew:1999nd}\\ \hline
			$p\to e^++\pi ^-+\rho ^+$ & 0 &   \\ \hline
			$p\to e^++\pi ^-+K^+$ & 0 & $75^*$ \cite{McGrew:1999nd} \\ \hline
			$p\to e^++2 \pi ^0$ & 0 & 147 \cite{McGrew:1999nd} \\ \hline
			$p\to e^+ +\pi ^0+\eta$ & 0 &   \\ \hline
			$p\to e^++\pi ^0+\rho ^0$ & 0 &  \\ \hline
			$p\to e^++\pi ^0+\omega$ & 0 &   \\ \hline
			$p\to e^++\pi ^0+K^0$ & 0 &   \\ \hline
			$p\to \mu ^-+2 \pi ^+$ & 2 & 17 \cite{Berger:1991fa} ($133^*$ \cite{McGrew:1999nd}) \\ \hline
			$p\to \mu ^-+K^++\pi ^+$ & 2 & 245 \cite{McGrew:1999nd} \\ \hline
			$p\to \mu ^++2 \gamma$ & 0 & $529^*$ \cite{McGrew:1999nd}\\ \hline
			$p\to \mu ^++\pi ^-+\pi ^+$ & 0 & 133 \cite{McGrew:1999nd} \\ \hline
			$p\to \mu ^++K^-+\pi ^+$ & 0 & $245^*$ \cite{McGrew:1999nd}\\ \hline
			$p\to \mu ^++\pi ^-+K^+$ & 0 & $245^*$ \cite{McGrew:1999nd}\\ \hline
			$p\to \mu ^++2 \pi ^0$ & 0 & 101 \cite{McGrew:1999nd} \\ \hline
			$p\to \mu ^+ +\pi ^0+\eta$ & 0 &  \\ \hline
			$p\to \mu ^++\pi ^0+K^0$ & 0 &   \\ \hline
			$p\to \nu+\pi ^++\pi ^0$ & 0,2 &   \\ \hline
			$p\to \nu +\pi ^++\eta$ & 0,2 &   \\ \hline
			$p\to \nu+\pi ^++\rho ^0$ & 0,2 &  \\ \hline
			$p\to \nu+\pi ^++\omega$ & 0,2 &   \\ \hline
			$p\to \nu+\pi ^++K^0$ & 0,2 &   \\ \hline
			$p\to \nu+\rho ^++\pi ^0$ & 0,2 &   \\ \hline
			$p\to \nu+K^++\pi ^0$ & 0,2 &   \\ \hline
		\end{tabular}
	\end{center}
	\caption{Same as Tab.~\ref{tab:two-body_limits} but for three-body proton decays. We have not included the (anyways unconstrained) $N\to A+B+\gamma$ modes since they should be suppressed compared to the $N\to A+B$ channels. We have however kept $N\to A +2\gamma$ because limits on this mode exist and one can imagine models with $\Gamma (N\to A +2\gamma)> \Gamma (N\to A +\gamma)$, e.g.~via a decay chain $N\to A + (J\to\gamma\gamma)$ with an axion-like particle $J$.
		\label{tab:three-body_proton_limits}}
\end{table}

\begin{table}[htbp!]
	\begin{center}
		\begin{tabular}{lcr}
			Channel & $|\Delta(B-L)|$ & $\frac{\Gamma^{-1}}{\unit[10^{30}]{yr}}$ \\ \hline\hline
			$n\to \nu+e^-+e^+$ & 0,2 & 257 \cite{McGrew:1999nd} \\ \hline
			$n\to \nu+e^-+\mu ^+$ & 0,2 & 83 \cite{McGrew:1999nd} \\ \hline
			$n\to \nu+e^++\mu ^-$ & 0,2 & $83^*$ \cite{McGrew:1999nd} \\ \hline
			$n\to \nu+\mu ^-+\mu ^+$ & 0,2 & 79 \cite{McGrew:1999nd} \\ \hline
			$n\to 3 \nu$ & 0,2,4 & $0.58$ \cite{Araki:2005jt} \\ \hline
			$n\to e^-+\pi ^++\pi ^0$ & 2 & 29 \cite{Berger:1991fa} ($52^*$ \cite{McGrew:1999nd})   \\ \hline
			$n\to e^-+\pi ^++\eta $ & 2 &   \\ \hline
			$n\to e^-+\pi ^++\rho ^0$ & 2 &  \\ \hline
			$n\to e^-+\pi ^++\omega$ & 2 &   \\ \hline
			$n\to e^-+\pi ^++K^0$ & 2 &   \\ \hline
			$n\to e^-+\rho ^++\pi ^0$ & 2 &   \\ \hline
			$n\to e^-+K^++\pi ^0$ & 2 &   \\ \hline
			$n\to e^++\pi ^-+\pi ^0$ & 0 & 52 \cite{McGrew:1999nd} \\ \hline
			$n\to e^++\pi ^-+\eta $ & 0 &   \\ \hline
			$n\to e^++\pi ^-+\rho ^0$ & 0 &   \\ \hline
			$n\to e^++\pi ^-+\omega$ & 0 &   \\ \hline
			$n\to e^++\pi ^-+K^0$ & 0 & 18 \cite{Berger:1990kg} \\ \hline
			$n\to e^++\rho ^-+\pi ^0$ & 0 &   \\ \hline
			$n\to e^++K^-+\pi ^0$ & 0 &   \\ \hline
			$n\to \mu ^-+\pi ^++\pi ^0$ & 2 & 34 \cite{Berger:1991fa} ($74^*$ \cite{McGrew:1999nd})  \\ \hline
			$n\to \mu ^- +\pi ^++\eta$ & 2 &   \\ \hline
			$n\to \mu ^-+\pi ^++K^0$ & 2 &   \\ \hline
			$n\to \mu ^-+K^++\pi ^0$ & 2 &   \\ \hline
			$n\to \mu ^++\pi ^-+\pi ^0$ & 0 & 74 \cite{McGrew:1999nd} \\ \hline
			$n\to \mu ^+ +\pi ^-+\eta$ & 0 &   \\ \hline
			$n\to \mu ^++\pi ^-+K^0$ & 0 &   \\ \hline
			$n\to \mu ^++K^-+\pi ^0$ & 0 &   \\ \hline
			$n\to \nu+2 \gamma$ & 0,2 & $219$ \cite{McGrew:1999nd}  \\ \hline
			$n\to \nu+\pi ^-+\pi ^+$ & 0,2 &   \\ \hline
			$n\to \nu+\rho ^-+\pi ^+$ & 0,2 &   \\ \hline
			$n\to \nu+K^-+\pi ^+$ & 0,2 &   \\ \hline
			$n\to \nu+\pi ^-+\rho ^+$ & 0,2 &   \\ \hline
			$n\to \nu+\pi ^-+K^+$ & 0,2 &   \\ \hline
			$n\to \nu+2 \pi ^0$ & 0,2 &   \\ \hline
			$n\to \nu +\pi ^0+\eta$ & 0,2 &   \\ \hline
			$n\to \nu+\pi ^0+\rho ^0$ & 0,2 &   \\ \hline
			$n\to \nu+\pi ^0+\omega$ & 0,2 &   \\ \hline
			$n\to \nu+\pi ^0+K^0$ & 0,2 &   \\ \hline
		\end{tabular}
	\end{center}
	\caption{Same as Tab.~\ref{tab:three-body_proton_limits}, but for neutron decays.
		\label{tab:three-body_neutron_limits}}
\end{table}

\section{Inclusive nucleon decay searches for \texorpdfstring{$\Delta B = 1$}{Delta B = 1} processes}
\label{sec:inclusive_searches}

As we have shown in Sec.~\ref{sec:theory}, without focusing on specific models many $\Delta B=1$ operators induce nucleon decays with a multi-body final state. More so, these channels could be dominant and often the simpler two-body decay modes are completely forbidden, e.g.~by a lepton number or flavor symmetry. It is therefore necessary to broaden the nucleon decay searches, which have primarily been focused on two-body channels, in order to also cover multi-body channels. While this can be done by performing exclusive searches looking at specific multi-body channels, an exhaustive search program quickly becomes impractical as the more complex modes are considered. Hence, fully model-independent as well as inclusive $N \rightarrow X + \text{anything}$ channel searches that are less dependent on the number and details of final-state particles are particularly important. We outline such searches below.

\subsection{Model-independent and invisible mode searches}

Model-independent searches allow us to probe all possible nucleon decay channels simultaneously and are not constrained to any specific final-state particle.
The most stringent model-independent limits come from nuclear de-excitation emission searches. Such limits apply to all seemingly unconstrained channels described in Tabs.~\ref{tab:two-body_limits},~\ref{tab:three-body_proton_limits}, and~\ref{tab:three-body_neutron_limits}.

Model-independent searches for neutron decays also constitute the main method to study invisible decays that leave no other trace in the detector, such as $n \rightarrow \text{invisible}$, where the invisible final state could consist of an arbitrary combination of neutrinos and antineutrinos. Invisible modes decaying to SM particles, such as $n \rightarrow 3 \nu$, could become significant in models based on extra dimensions~(e.g.~\cite{Dvali:1999hn,Mohapatra:2002ug}) or partially unified Pati--Salam-type constructions~\cite{Pati:1973rp}. Proposals for invisible decays to more exotic states, such as dark fermions~\cite{Fornal:2018eol,Barducci:2018rlx} or unparticles~\cite{He:2008rv}, have been also suggested.
Applying these searches to processes like $p \rightarrow 3\nu$ can test electric-charge conservation.

\subsubsection{Isotope fission products}

A model-independent search for nucleon decays can be performed by analyzing fission isotope products due to nucleon decay within an element~\cite{Primakoff:1981sx}. If the fission fragment is stable (unstable), a geochemical (radiogenic/radiochemical) search method can be used.

In the radiogenic/radiochemical approach  one looks for radioactive remnants due to nucleon decay within some favorable and abundant isotope, such as $^{58}\text{Ni}\rightarrow$ $^{57}$Co~\cite{Steinberg:1976gc} or $^{39}\text{K}\rightarrow (^{38}\text{Ar},{}^{38}\text{K})\rightarrow {}^{37}\text{Ar}$~\cite{Fireman:1979xr}. These experiments must be placed deep underground to suppress background isotope production due to cosmic ray interactions. Further, as radioactive decay happens during experimental observation large quantities of material are required. Typically, these experiments are not easily scalable, with limits on nucleon decay lifetimes of $\tau_p \sim \unit[10^{26\text{--}27}]{yr}$ already requiring ton-scale fiducial mass. 

Due to their ultra-low keV-level thresholds and well-understood backgrounds, large DM experiments could also constitute efficient detectors for radiogenic nucleon decay searches. Using $\sim \unit[7]{kg}$ of fiducial mass and $\sim \unit[6]{kg\cdot yr}$ of exposure, a limit of $\tau_p \sim \unit[10^{24}]{yr}$ was obtained with the DAMA liquid Xenon scintillator~\cite{Bernabei:2000xp}. A future large-scale Xenon experiment with fiducial mass of $\mathcal{O}(10\text{--}100)$ tons, such as DARWIN~\cite{Aalbers:2016jon}, might thus achieve an improved sensitivity reach by several orders of magnitude. 

A variation on the radiogenic approach is to search for appearance  of  fragments from   spontaneous fission of a large nucleus into several sizable components. Here, using $\unit[7]{kg \cdot hr}$ exposure of $^{232}$Th a limit of $\tau_p \sim \unit[10^{21}]{yr}$ was set in Ref.~\cite{Flerov:1958zz}. However, using this method effectively requires good background understanding and large quantities of very heavy elements.

In the geochemical approach one searches for stable nuclear residue associated with nucleon decays that have accumulated over billions of years within naturally abundant ore.
This approach has been employed for double beta-decay studies, especially when the resulting isotope is a noble gas. When applied to nucleon decay by considering processes such as ${}^{23}\text{Ne}\rightarrow {}^{22}\text{Ne}$, ${}^{39}\text{K}\rightarrow {}^{38}\text{Ar}$, and ${}^{133}\text{Cs}\rightarrow {}^{132}\text{Xe}$, a nucleon decay lifetime sensitivity of $\sim \unit[10^{23}]{yr}$ can be achieved~\cite{Rosen:1975ch}, assuming
spectroscopic resolution of
1~part in $10^8$ (\unit[10]{ppb}).
An improved lifetime limit of $\sim \unit[10^{25}]{yr}$ was achieved in Ref.~\cite{Steinberg:1977it} by considering beta decays in ${}	^{130}\text{Te}\rightarrow {}^{129}\text{Sb}\rightarrow {}^{129}\text{Xe}$, measured double beta-decay lifetime and known abundances as well as atmospheric contamination of various xenon isotopes in telluride ore.
Due to the very long lifetime of the source material, these searches are not limited by fiducial mass as radiochemical searches are. These searches require precise impurity identification. 
While purification levels of material, such as ${}^{130}\text{Te}$ crystals produced for double beta-decay studies, in current experiments has greatly improved in recent years and approaches 1 part in $10^{14}$~\cite{Arnaboldi:2010fj}, obtaining accurate understanding of accumulated background contributions  is highly non-trivial.

While we envision that some improvement on such searches is feasible, it is difficult for these searches to compete with nuclear de-excitation emission searches.

\subsubsection{Nuclear de-excitation emission}
\label{sec:deexcitation}

Aside from analyzing fission fragments, another general and model-independent signature of nucleon decay is nuclear de-excitation emission. 
Every nucleon decay results in a hole within the host nucleus, e.g.~in oxygen $^{16}$O (as relevant for WC-based detectors) or carbon $^{12}$C (as relevant for scintillator-based detectors). 
Such decays from an inner nuclear shell typically leave the  residual nucleus in an excited state, which subsequently de-excites by emission of secondary $(n, p, \alpha, \gamma)$ particles and also possibly further via $\beta$-decay of the residual radioactive nuclei. Emission steps can be identified with nuclear-shell models and energetic considerations. The de-excitation particles accompanying nucleon decay could be searched for as a model-independent signature~\cite{Totsuka:1986zn,Ejiri:1993rh,Kamyshkov:2002wp,Boyd:2003xk}. Such signals, however, could be plagued by typical radioactive background within experiments as well as neutrino interactions that result in similarly excited nuclear states~\cite{Wan:2019xnl}.

For WC experiments, de-excitation $\gamma$'s are especially important for these searches. De-excitation $\gamma$ emission with energies of a few MeV is expected with a large branching ratio (e.g.~$44\%$ for  $\unit[6.18]{MeV}$ $\gamma$-ray for $^{15}$O~\cite{Ejiri:1993rh}). However, this energy range is plagued by backgrounds associated with solar neutrinos and cosmogenics. Placing the experiment very deep underground (e.g.~6000 meter water equivalent for SNO+) and using a directional cut to suppress solar neutrino contributions, the SNO+ scintillator experiment was able to achieve a limit on invisible proton decay of
\begin{align}
    \tau_p^\text{inv.}  > \unit[0.36\times 10^{30}]{yr}
\end{align}
with only $\unit[0.9]{kton}$ of water and $\unit[0.58]{kton}\cdot$yr exposure obtained during its water-filled WC phase~\cite{Anderson:2018byx}.

Large WC experiments that have significant exposure can take advantage of a relatively clean signature from the higher energy $\gamma$-rays, which however have suppressed branching ratios. In particular, search for $\gamma$-rays in the 19--$\unit[50]{MeV}$ energy range has been used to set a lower partial-lifetime limit associated with $n \rightarrow 3\nu$ of $\tau/\text{Br} \sim \unit[2 \times 10^{31}]{yr}$ by the Kamiokande experiment using $\sim \unit[8.5]{kton}\cdot$yr exposure~\cite{Suzuki:1993zp}. With the associated cumulative branching ratio for such energetic $\gamma$'s of $(1.4 \pm 0.7) \times 10^{-4}$~\cite{Kamyshkov:2002wp}, the resulting lifetime limit is $\tau \sim \unit[5 \times 10^{26}]{yr}$~\cite{Suzuki:1993zp}. With over $\unit[370]{kton}\cdot$yr of exposure available for SK, a limit of order $\unit[10^{28}]{yr}$ can  be achieved using this method.

A coincidence multi-event signature associated with several spatially and temporally correlated secondary particles from de-excitation can be employed to differentiate signal from background (see discussion in Ref.~\cite{Kamyshkov:2002wp}). Such low-energy MeV-level signals associated with de-excitation can be particularly advantageous for scintillator-based experiments, as considered by Borexino~\cite{Back:2003wj} and KamLAND~\cite{Araki:2005jt}. In particular, a limit on invisible nucleon decay of
\begin{align} \label{eq:invnlim}
\tau_n^\text{inv.} > \unit[0.58\times 10^{30}]{yr}
\end{align}
was obtained with the $\unit[3.2]{kton}$ KamLAND scintillator detector and $\unit[0.84]{kton}\cdot$yr of data~\cite{Araki:2005jt}.

In WC-based experiments, a de-excitation proton or an $\alpha$ would be below the Cherenkov threshold and thus invisible. A kicked-out neutron quickly thermalizes and diffuses, subsequently being captured on a hydrogen atom with an accompanying emission of a low-energy $\unit[2.2]{MeV}$ $\gamma$-ray. A low-efficiency neutron tagging is currently being used in SK to gather this signal~\cite{Super-Kamiokande:2015xra}, which is below the energy threshold of SK. The planned SK upgrade involving gadolinium doping~\cite{Beacom:2003nk} will allow us to detect neutrons with high efficiency due to $\sim \unit[8]{MeV}$ $\gamma$-ray emission accompanying neutron capture on gadolinium. Since de-excitations will generally result in neutrons,\footnote{Taking into account both single-step and multi-step emission, neutrons are expected to appear with a combined branching ratio of $\sim 50\%$ from de-excitation of $^{15}$O associated with neutron decays from an $s_{1/2}$ state of $^{16}$O~\cite{Kamyshkov:2002wp}.} they could be beneficial for WC searches.
However, without additional coincidence signatures the signal-background discrimination is non-trivial as neutrons also generically appear from neutrino interactions such as inverse $\beta$-decay $\overline{\nu}_e + p \to n + e^+$. Previously, the importance of detecting neutrons associated with proton decay has been emphasized for heavy water (D$_2$O) searches (e.g.~in SNO)~\cite{Tretyak:2001fv}. In particular, the proton decay search $d \to n + ?$ requires less stringent implicit hypotheses regarding the stability of the resulting daughter nuclei and is hence even less model dependent than some of the other searches discussed in this section.

Coincidence tagging of de-excitation emission together with some visible final state can provide additional discrimination of nucleon decay signal versus backgrounds, as first suggested by Totsuka for SK~\cite{Totsuka:1986zn}. However, this is often difficult to utilize in practice. For example, in WC searches the nearly prompt emission of a low-energy de-excitation $\gamma$ could be hard to resolve when super-imposed with an energetic Cherenkov ring e.g.~from a final state $e^{\pm}$ or $\mu^{\pm}$. A time separation between the energy depositions, for example in $p\to\nu K^+$ due to the $\unit[12]{ns}$ kaon lifetime, is then beneficial and can be used to suppress background -- as is done in SK analyses~\cite{Hayato:1999az,Kobayashi:2005ut,Kobayashi:2006gb,Abe:2014mwa}.

Future detectors such as JUNO, DUNE, and HK are expected to improve on these important channels. While we are not aware of dedicated sensitivity studies, the JUNO liquid-scintillator detector, which is analogous to KamLAND but with a 20 times larger size (\unit[20]{kt}), could be expected to test invisible-nucleon decays beyond $\unit[10^{31}]{yr}$ lifetime.  Further, with low detection thresholds and sizable \unit[40]{kt} mass, DUNE could also be very promising. To our knowledge no dedicated studies for de-excitation emission associated with nucleon decay from argon, as would be relevant for DUNE, have been performed thus far.

Despite being derived as limits on invisible nucleon decay, the discussed limits can be interpreted as the best current model-independent limits on the \emph{total} nucleon lifetime and are likely applicable to all otherwise unconstrained modes in Tabs.~\ref{tab:two-body_limits}--\ref{tab:three-body_neutron_limits}. We note that there is still \emph{some} implicit dependence on the actual nucleon decay mode. For example, decay modes with highly energetic pions can potentially destroy the daughter nucleus rather than just leaving a hole, but we envision that such events would be captured with other searches. 

Finally, we comment that the invisible searches presented here could also be interpreted as tests of forbidden nuclear transitions, as was done in Ref.~\cite{Suzuki:1993zp}.
 
\subsection{\texorpdfstring{$N\to X +$}{N to X+}anything}
\label{sec:NtoXplusanything}

Above we have discussed nucleon decay searches that are nearly model independent and only make use of the properties of the daughter nucleus that is created after nucleon decays within a nucleus. The current most sensitive searches in this category are from de-excitation emission, with resulting lifetime limits approaching $\unit[10^{30}]{yr}$. This is more than four orders of magnitude below the best exclusive search limits (Tab.~\ref{tab:two-body_limits}) and one or two orders of magnitude below the predominant majority of other existing exclusive searches (e.g.~Tabs.~\ref{tab:three-body_proton_limits} and~\ref{tab:three-body_neutron_limits}).
In an effort to keep searches as model independent as possible while utilizing the excellent particle identification of exclusive searches, we now discuss \emph{inclusive} nucleon decay searches, which correspond to $N\to X+\text{anything}$ with a light SM particle $X$ of unknown energy.
Without knowledge of the underlying energy distribution these are in general background-dominated searches.

\subsubsection{\texorpdfstring{$N \to (e^{\pm}, \mu^{\pm}, \pi^{\pm}, K^{\pm}, \rho^{\pm}, K^{\ast, \pm})~+$}{N to charged particle +}~anything}

In general, from size and exposure considerations, WC experiments are expected to dominate inclusive searches with a charged particle. However, the primary charged particle of interest could be invisible in WC experiments due to its Cherenkov threshold. Hence, additional detection efficiency can be expected in experiments with low-energy thresholds and where they are clearly visible, such as scintillators, especially for particles with a higher mass and a larger Cherenkov threshold.
 
From electric-charge conservation it is clear that any proton decay will eventually result in at least one positron, albeit potentially space-time-delayed if it originated from the decay of a heavier charged final-state particle (e.g.~muon or pion). 
Hence, $p \rightarrow e^{\pm} + \text{anything}$ is an important general inclusive search.
Since the positron will almost always have its momentum above the Cherenkov threshold, unless the number of final-state particles is very high, a particularly promising sensitivity  for this  general inclusive search is achievable from a re-analysis of a one-ring $e$-like data sample in SK and subsequently HK. The current best lifetime limit on the $N \rightarrow e^+ + \text{anything}$ search is $\unit[0.6\times 10^{30}]{yr}$, from a 1979 analysis of Ref.~\cite{Learned:1979gp}. We expect that SK and later HK could significantly improve on this result, due to their large detector volume and exposure. An estimate of such a limit can be obtained from the recent SK search that placed a limit of
$\unit[170\times 10^{30}]{yr}$~\cite{Takhistov:2014pfw} on the
$p\to e^+\nu\nu$ channel, which bears close similarity to the inclusive $p \to e^+ + \text{anything}$ mode.   Lowering the positron-momentum threshold employed in that analysis and varying the energy distribution, or even adapting a pure counting analysis, is expected to yield an inclusive $N \rightarrow e^\pm + \text{anything}$ mode limit of similar order with available SK data, i.e.~$\unit[\mathcal{O}(100)\times 10^{30}]{yr}$.

In the case of inclusive  $N \rightarrow \mu^\pm + \text{anything}$ channel with a muon, if the muon has enough momentum it leads to a $\mu$-like ring in WC detectors. The current limit on this search is $\tau (N\to\mu^++\text{anything}) > \unit[12\times 10^{30}]{yr}$~\cite{Tanabashi:2018oca}, converted from a 1981 Homestake experiment analysis of Ref.~\cite{Cherry:1981uq} that used a $\unit[0.3]{kton}$ WC detector. We again expect SK and future HK to improve on this limit. Similar to the inclusive search with an electron, the recent SK search of $p\to \mu^+\nu\nu$ that resulted in a lifetime limit of $\unit[220\times 10^{30}]{yr}$~\cite{Takhistov:2014pfw} serves as a good indication of the potential limit on $N \rightarrow \mu^\pm + \text{anything}$.

Inclusive searches for charged mesons are feasible as well, even though the mesons will eventually decay into charged leptons and  thus in principle will be   covered by the searches outlined above. Nevertheless, designated charged-meson searches can provide additional useful information. In WC detectors the charged pion $\pi^\pm$ will look similar to a $\mu^\pm$, albeit with a somewhat higher Cherenkov threshold, so the resulting limit can be expected to be of similar order.
An estimate of $p \to \pi^{\pm} + $anything can be obtained from the SK search for the two-body decay $p \to \nu + \pi^+$  that yielded a limit of $390 \times 10^{30}$~yr~\cite{Abe:2013lua}. Unlike this two-body decay, multi-body decays will result in a smeared pion momentum -- causing an additional detection-efficiency penalty when the pion is below Cherenkov threshold.

Charged mesons more massive than the pion (i.e.~$K^{\pm}, \rho^{\pm}, K^{\ast, \pm})$ are always below Cherenkov threshold when originating from decays of a single nucleon.
Thus, in WC experiments such mesons can only be identified by their decay products. Due to their low momentum and short lifetime, they effectively decay at rest. There is thus not much dependence of the associated inclusive limits on momentum of the heavy meson and if the decay is two body or multi-body. Hence, limits from exclusive searches with such mesons and other states being invisible, such as $\tau(p \to \nu + K^+) > \unit[5900 \times 10^{30}]{yr}$~\cite{Abe:2014mwa} and $\tau(p \to \nu + K^{\ast, +}) > \unit[130 \times 10^{30}]{yr}$~\cite{Scott:2007} from SK as well as $\tau(p \to \nu + \rho^+) > \unit[162 \times 10^{30}]{yr}$ from IMB~\cite{McGrew:1999nd}, can be readily re-interpreted as inclusive limits on $p \to (K^{\pm}, \rho^{\pm}, K^{\ast, \pm}) + $anything processes.

We expect that SK can improve by more than an order over the existing IMB limits and even further improvement can be achieved with future HK. Further, JUNO and DUNE are particularly promising for complementary studies of these modes since they allow for an identification of the low-energy heavy charged mesons that are below Cherenkov threshold, with a particularly good efficiency achievable for kaons.

\subsubsection{\texorpdfstring{$N\to (\pi^0,K^0,\eta,\rho, \omega, K^{\ast, 0})~+$}{N to neutral meson +}~anything}

Inclusive searches involving neutral mesons $M^0$ rely on the electromagnetically interacting daughter particles of $M^0$ for detection. For example, in the rest frame $\pi^0\to \gamma\gamma$ decay results in photon lines of energy $m_\pi/2$, clearly visible in WC detectors. For slow $\pi^0$ these are back to back, otherwise more collimated.
Already in the two-body nucleon decay the pion is not very energetic and carries energy below $m_N/2$. Hence, the inclusive limit $n \to \pi^0 + $anything is not particularly sensitive to pion momentum and one can estimate it to be of  similar order as the limit from existing SK search for $n\to\pi^0 \nu$ of $\unit[1100 \times 10^{30}]{yr}$~\cite{Abe:2013lua}.

The above considerations are even more applicable to neutral mesons heavier than pion (i.e.~$K^0,\eta,\rho^0, \omega, K^{\ast, 0}$), which can be approximated as decaying at rest.
Hence, the search limits of  $\tau(n \to \nu + K^0) > \unit[130 \times 10^{30}]{yr}$~\cite{Kobayashi:2005pe} from SK,  $\tau(n \to \nu + \eta) > \unit[158 \times 10^{30}]{yr}$~\cite{McGrew:1999nd} from IMB, $\tau(n \to \nu + \rho^0) > \unit[19 \times 10^{30}]{yr}$~\cite{Seidel:1988ut} from IMB, $\tau(n \to \nu + \omega) > \unit[108 \times 10^{30}]{yr}$~\cite{McGrew:1999nd} from IMB and $\tau(n \to \nu + K^{\ast, 0}) > \unit[78 \times 10^{30}]{yr}$~\cite{McGrew:1999nd} from IMB 
can provide an estimate of magnitude for sensitivity to such inclusive modes.
 
As before, we expect that a re-analysis of the above modes with the SK data-set can improve IMB's limits by more than an order of magnitude and even better limits can be expected from HK.
	
\subsubsection{\texorpdfstring{$N\to \gamma~+$}{N to gamma +}~anything}

Since we are envisioning multi-body nucleon decays, the emission of a photon is typically suppressed compared to the same channel without a photon. It is still useful to perform such an analysis for final states that are otherwise difficult to detect, such as $n\to\gamma+\text{neutrinos}$. As emphasized in Ref.~\cite{Glicenstein:1997pv}, neutron decay into neutrinos corresponds to the sudden disappearance of the neutron's magnetic moment  and should thus lead to electromagnetic radiation. From a particle-physics perspective this effect can be readily obtained from attaching a photon to the initial quarks in any diagram leading to $n\to \text{neutrinos}$, in analogy to radiative $\tau$ decays~\cite{Laursen:1983sm}.
Compared to low-energy de-excitation photons, these photons can reach energies of up to $m_N/2$ and thus provide a far cleaner signature. However, with the probability that a photon emitted from this process has energy $> \unit[100]{MeV}$ being only $\sim 5 \times 10^{-5}$~\cite{Glicenstein:1997pv}, it is unlikely that such a search will be more sensitive than the $n \to$ invisible limit of Eq.~\eqref{eq:invnlim}. Further, the photon spectrum here depends on the details of the final state and number of emitted neutrinos, implying that obtained limits from such a search are less general than Eq.~\eqref{eq:invnlim}.

Nevertheless, this search would provide some complimentary to the low-energy $\gamma$ searches from de-excitation and is also close to the inclusive search on $e^{\pm}$, as both result in an $e$-like ring in WC experiments.
 
\subsubsection{\texorpdfstring{$N\to \nu~+$}{N to neutrino +}~anything} 
\label{sec:invisible}

As neutrinos escape the detector before interacting, this search is effectively invisible if nucleon decays are considered to occur within the experimental fiducial volume. However, it has been suggested that by considering such decays on macroscopic Earth-sized scales the additional cumulative neutrino flux from decays can be significant and observable. By attributing the experimentally observed $\nu_{\mu}/\nu_e$ fluxes to nucleon decays in Earth a limit of $\sim \unit[10^{26}]{yr}$ was estimated~\cite{Learned:1979gp}. However, a further reanalysis by the Fr{\'e}jus experiment using $\unit[2]{kton}\times\unit{yr}$ of data yielded a lower limit of $\sim \unit[10^{25}]{yr}$~\cite{Berger:1991fa} (see their Sec.~7 for flux estimation). Since SK has already collected over $\unit[370]{kton} \times \unit{yr}$ of data (e.g.~\cite{Sussman:2018ylo}), we anticipate that an improved sensitivity of up to $\sim \unit[10^{28}]{yr}$ could be achieved by re-analysis of this channel in SK using the full data set and even further improved upon in future HK. We note that exact search details depend on the assumptions about the flavor of  emitted neutrinos.

\section{Processes with \texorpdfstring{$\Delta B > 1$}{Delta B > 1}}
\label{sec:higher_B}

So far we have focused on processes violating baryon number by one unit, $\Delta B =1$, which are described by the lowest-dimension $\Delta B\neq 0$ SMEFT operators. 
However, nucleon decays also constitute sensitive probes of dimension $d\gg 6$ operators beyond $\Delta B = 1$.
Such ``multi-nucleon decays'' with $\Delta B >1$ can be treated in complete analogy to the processes discussed in previous sections. The main difference is that $\Delta B >1$ processes have more available energy, which allows for the on-shell production of tau leptons~\cite{Bryman:2014tta} as well as heavier mesons such as $D$ and $\phi$, thus increasing the number of possible final states.
Appearance of visible heavier charged mesons such as $K^{\pm}$, which are always below Cherenkov threshold when originating from single nucleon decays, also becomes possible.
Only a few channels out of all possible kinematically allowed two-body $\Delta B =2$ di-nucleon decays have been experimentally studied thus far (17 out of 118), as we summarize in Tabs.~\ref{tab:two-body_dinucleon_limits1}--\ref{tab:two-body_dinucleon_limits3}. Even more limited are searches for
$\Delta B >2$ processes, as we briefly comment on in Sec.~\ref{sec:DeltaBabove2}.

\begin{table}[htbp!]
	\begin{center}
		\begin{tabular}{lcr}
			Channel & $|\Delta(B-L)|$ & $\frac{\Gamma^{-1}}{\unit[10^{30}]{yr}}$ \\ \hline\hline
			$pp\to e^++e^+$ & 0 &  $4200$ \cite{Sussman:2018ylo}\\ \hline
			$pp\to \mu^++\mu^+$ & 0 &  $4400$ \cite{Sussman:2018ylo}\\ \hline
			$pp\to e^++\mu^+$ & 0 &  $4400$ \cite{Sussman:2018ylo}\\ \hline
			$pp\to e^++\tau^+$ & 0 &     \\ \hline
			$pp\to \pi^++\pi^+$ & 2 &  $72$ \cite{Gustafson:2015qyo}\\ \hline
			$pp\to \pi^++\rho^+$ & 2 &     \\ \hline
			$pp\to \pi^++K^+$ & 2 &     \\ \hline
			$pp\to \pi^++K^{*,+}$ & 2 &     \\ \hline
			$pp\to \rho^++\rho^+$ & 2 &     \\ \hline
			$pp\to \rho^++K^+$ & 2 &     \\ \hline
			$pp\to \rho^++K^{*,+}$ & 2 &     \\ \hline
			$pp\to K^++K^+$ & 2 &  $170$ \cite{Litos:2014fxa}\\ \hline
			$pp\to K^++K^{*,+}$ & 2 &     \\ \hline
			$pp\to K^{*,+}+K^{*,+}$ & 2 &     \\ \hline
			\hline
			$nn\to e^++e^-$ & 2 &  $4200$ \cite{Sussman:2018ylo}\\ \hline
			$nn\to e^++\mu^-$ & 2 &  $4400$ \cite{Sussman:2018ylo}\\ \hline
			$nn\to \mu^++e^-$ & 2 &  $4400$ \cite{Sussman:2018ylo}\\ \hline
			$nn\to \mu^++\mu^-$ & 2 &  $4400$ \cite{Sussman:2018ylo}\\ \hline
			$nn\to e^++\tau^-$ & 2 &     \\ \hline
			$nn\to \tau^++e^-$ & 2 &     \\ \hline
			$nn\to 2 \nu$ & 0,2,4 &  $1.4$ \cite{Araki:2005jt}\\ \hline
			$nn\to 2 \gamma$ & 2 &  $4100$ \cite{Sussman:2018ylo}\\ \hline
			$nn\to \gamma +\pi ^0 $ & 2 &     \\ \hline
			$nn\to \gamma +\eta$ & 2 &     \\ \hline
			$nn\to \gamma +\rho ^0 $ & 2 &     \\ \hline
			$nn\to \gamma +\omega  $ & 2 &     \\ \hline
			$nn\to \gamma +\eta '$ & 2 &     \\ \hline
			$nn\to \gamma +K^0 $ & 2 &     \\ \hline
			$nn\to \gamma +K^{*,0}$ & 2 &     \\ \hline
			$nn\to \gamma +D^0 $ & 2 &     \\ \hline
			$nn\to \gamma +\phi$ & 2 &     \\ \hline
			$nn\to \pi ^-+\pi ^+ $ & 2 &  $0.7$ \cite{Berger:1991fa} ($72^*$ \cite{Gustafson:2015qyo})\\ \hline
			$nn\to \pi ^++\rho ^- $ & 2 &     \\ \hline
			$nn\to K^-+\pi ^+ $ & 2 &     \\ \hline
			$nn\to K^{*,-}+\pi ^+$ & 2 &     \\ \hline
			$nn\to \pi ^-+\rho ^+ $ & 2 &     \\ \hline
			$nn\to K^++\pi ^- $ & 2 &     \\ \hline
			$nn\to K^{*,+}+\pi ^-$ & 2 &     \\ \hline
			$nn\to 2 \pi ^0 $ & 2 &  $404$ \cite{Gustafson:2015qyo}\\ \hline
			$nn\to \eta +\pi ^0$ & 2 &     \\ \hline
			$nn\to \pi ^0+\rho ^0$ & 2 &     \\ \hline
			$nn\to \pi ^0+\omega$ & 2 &     \\ \hline
			$nn\to \eta '+\pi ^0 $ & 2 &     \\ \hline
			$nn\to K^0+\pi ^0 $ & 2 &     \\ \hline
			$nn\to K^{*,0}+\pi ^0 $ & 2 &     \\ \hline
		\end{tabular}
	\end{center}
	\caption{Same as Tab.~\ref{tab:two-body_limits}, but listing all kinematically allowed two-body \emph{di-nucleon} $\Delta B=2$  decays with $90\%$~C.L.~upper limit on the partial decay width $\Gamma$. 
		\label{tab:two-body_dinucleon_limits1}}
\end{table}

\begin{table}[htbp!]
	\begin{center}
		\begin{tabular}{lcr}
			Channel & $|\Delta(B-L)|$ & $\frac{\Gamma^{-1}}{\unit[10^{30}]{yr}}$ \\ \hline\hline
			$nn\to \pi ^0+\phi$ & 2 &     \\ \hline
			$nn\to 2 \eta$ & 2 &     \\ \hline
			$nn\to \eta +\rho ^0 $ & 2 &     \\ \hline
			$nn\to \eta +\omega $ & 2 &     \\ \hline
			$nn\to \eta +\eta '$ & 2 &     \\ \hline
			$nn\to \eta +K^0 $ & 2 &     \\ \hline
			$nn\to \eta +K^{*,0} $ & 2 &     \\ \hline
			$nn\to \eta +\phi $ & 2 &     \\ \hline
			$nn\to 2 \rho ^0$ & 2 &     \\ \hline
			$nn\to \rho ^0+\omega  $ & 2 &     \\ \hline
			$nn\to \eta '+\rho ^0 $ & 2 &     \\ \hline
			$nn\to K^0+\rho ^0 $ & 2 &     \\ \hline
			$nn\to K^{*,0}+\rho ^0$ & 2 &     \\ \hline
			$nn\to \rho ^0+\phi$ & 2 &     \\ \hline
			$nn\to \rho ^-+\rho ^+$ & 2 &     \\ \hline
			$nn\to K^++\rho ^-$ & 2 &     \\ \hline
			$nn\to K^{*,+}+\rho ^- $ & 2 &     \\ \hline
			$nn\to K^-+\rho ^+ $ & 2 &     \\ \hline
			$nn\to K^{*,-}+\rho ^+$ & 2 &     \\ \hline
			$nn\to 2 \omega $ & 2 &     \\ \hline
			$nn\to \eta '+\omega $ & 2 &     \\ \hline
			$nn\to K^0+\omega $ & 2 &     \\ \hline
			$nn\to K^{*,0}+\omega$ & 2 &     \\ \hline
			$nn\to \omega +\phi$ & 2 &     \\ \hline
			$nn\to \eta '+K^0 $ & 2 &     \\ \hline
			$nn\to \eta '+K^{*,0}$ & 2 &     \\ \hline
			$nn\to K^-+K^+$ & 2 &   $170^*$ \cite{Litos:2014fxa}  \\ \hline
			$nn\to K^++K^{*,-} $ & 2 &     \\ \hline
			$nn\to K^-+K^{*,+} $ & 2 &     \\ \hline
			$nn\to 2 K^0$ & 2 &     \\ \hline
			$nn\to K^{*,0}+K^0$ & 2 &     \\ \hline
			$nn\to K^0+\phi $ & 2 &     \\ \hline
			$nn\to 2 K^{*,0} $ & 2 &     \\ \hline
			$nn\to K^{*,-}+K^{*,+}$ & 2 &     \\ \hline
		\end{tabular}
	\end{center}
	\caption{Extension of Tab.~\ref{tab:two-body_dinucleon_limits1}. 
		\label{tab:two-body_dinucleon_limits2}}
\end{table}

\begin{table}[htbp!]
	\begin{center}
		\begin{tabular}{lcr}
			Channel & $|\Delta(B-L)|$ & $\frac{\Gamma^{-1}}{\unit[10^{30}]{yr}}$ \\ \hline\hline
			$pn\to e^++\nu$ & 0,2 &  $260$ \cite{Takhistov:2015fao}\\ \hline
			$pn\to \mu^++\nu$ & 0,2 &  $200$ \cite{Takhistov:2015fao}\\ \hline
			$pn\to \tau^++\nu$ & 0,2 &  $29$ \cite{Takhistov:2015fao}\\ \hline
			$pn\to  \gamma +\pi ^+ $ & 2 &     \\ \hline
			$pn\to \gamma +\rho ^+  $ & 2 &     \\ \hline
			$pn\to  \gamma +K^+ $ & 2 &     \\ \hline
			$pn\to  \gamma +K^{*,+} $ & 2 &     \\ \hline
			$pn\to  \gamma +D^+ $ & 2 &     \\ \hline
			$pn\to  \pi ^++\pi ^0 $ & 2 &  $170$ \cite{Gustafson:2015qyo}\\ \hline
			$pn\to \eta +\pi ^+ $ & 2 &     \\ \hline
			$pn\to  \pi ^++\rho ^0 $ & 2 &     \\ \hline
			$pn\to  \pi ^++\omega $ & 2 &     \\ \hline
			$pn\to  \eta '+\pi ^+ $ & 2 &     \\ \hline
			$pn\to  K^0+\pi ^+ $ & 2 &     \\ \hline
			$pn\to  K^{*,0}+\pi ^+ $ & 2 &     \\ \hline
			$pn\to  \pi ^++\phi $ & 2 &     \\ \hline
			$pn\to  \pi ^0+\rho ^+$ & 2 &     \\ \hline
			$pn\to K^++\pi ^0  $ & 2 &     \\ \hline
			$pn\to  K^{*,+}+\pi ^0$ & 2 &     \\ \hline
			$pn\to  \eta +\rho ^+ $ & 2 &     \\ \hline
			$pn\to  \eta +K^+ $ & 2 &     \\ \hline
			$pn\to  \eta +K^{*,+}$ & 2 &     \\ \hline
			$pn\to  \rho ^++\rho ^0 $ & 2 &     \\ \hline
			$pn\to  K^++\rho ^0 $ & 2 &     \\ \hline
			$pn\to  K^{*,+}+\rho ^0 $ & 2 &     \\ \hline
			$pn\to  \rho ^++\omega $ & 2 &     \\ \hline
			$pn\to  \eta '+\rho ^+$ & 2 &     \\ \hline
			$pn\to  K^0+\rho ^+$ & 2 &     \\ \hline
			$pn\to  K^{*,0}+\rho ^+ $ & 2 &     \\ \hline
			$pn\to  \rho ^++\phi $ & 2 &     \\ \hline
			$pn\to  K^++\omega  $ & 2 &     \\ \hline
			$pn\to  K^{*,+}+\omega$ & 2 &     \\ \hline
			$pn\to  \eta '+K^+$ & 2 &     \\ \hline
			$pn\to  \eta '+K^{*,+} $ & 2 &     \\ \hline
			$pn\to  K^++K^0$ & 2 &     \\ \hline
			$pn\to  K^++K^{*,0} $ & 2 &     \\ \hline
			$pn\to  K^++\phi  $ & 2 &     \\ \hline
			$pn\to  K^{*,+}+K^0 $ & 2 &     \\ \hline
			$pn\to  K^{*,+}+K^{*,0}$ & 2 &     \\ \hline
		\end{tabular}
	\end{center}
	\caption{Extension of Tabs.~\ref{tab:two-body_dinucleon_limits1} and~\ref{tab:two-body_dinucleon_limits2}.
		\label{tab:two-body_dinucleon_limits3}}
\end{table}

Let us reiterate an important point: $\Delta B >1$ processes do not need to be suppressed compared to $\Delta B=1$ nucleon decays. In fact, it is possible to completely forbid all $|\Delta B| < n$ processes by a global symmetry, making $\Delta B =n$ the dominant baryon number violating process~\cite{Weinberg:1979sa}. This already occurs within the SM, where baryon number is only broken by three units, so the proton remains stable.
Models in which $\Delta B =2$ processes dominate over $\Delta B=1$ have been discussed extensively in the literature (e.g.~\cite{Mohapatra:1980qe,Rao:1982gt,Nussinov:2001rb,Arnold:2012sd,Girmohanta:2019fsx}).
In addition to total baryon and lepton number as selection rules for allowed processes~\cite{Weinberg:1979sa} one can further consider lepton flavor~\cite{Heeck:2016xwg,Hambye:2017qix}, which is the only other global quantum number not violated in SM interactions.

\subsection{\texorpdfstring{$\Delta B=2$}{Delta B = 2}, \texorpdfstring{$\Delta L =0$}{Delta L = 0}}

 As illustrated in Fig.~\ref{fig:BL_grid}, $\Delta B =2$ processes start at $d=9$ and do not involve leptons. The corresponding operators, such as~$udd udd$, have been discussed at length in the literature~\cite{Mohapatra:1980qe,Kuo:1980ew,Ozer:1982qh,Rao:1982gt}. Experimentally, they have been probed through neutron--antineutron ($n$--$\overline{n}$) oscillations~\cite{Phillips:2014fgb} as well as deuteron decays~\cite{Oosterhof:2019dlo} and other di-nucleon decays such as $nn\to\pi\pi$. Taking into account the possible quark-flavor structure of these operators it is clear that they can induce a variety of $nn\to M M'$ processes, with $M$ and $M'$ being mesons or photons, and similarly $np, pp\to\text{mesons}$ decays. While more complicated multi-meson final states are also expected, calculations of their branching ratios have not been extensively performed. Experimentally, only a small subset of such possible two-body final states has been searched for, namely $nn\to 2\gamma$, $N N' \to \pi \pi$ and $N N'\to K K$, with the latter being motivated by SUSY~\cite{Barbieri:1985ty,Goity:1994dq}.
 
$\Delta B =2$ operators with $\Delta L =0$ involving leptons arise at $d= 13$ and should be suppressed compared to the $d=9$ operators unless they carry lepton flavor~\cite{Heeck:2016xwg,Hambye:2017qix}, leading for example to~$nn\to\mu^+e^-$ or $nn\to\mu^+\tau^-$. The tau modes have not been tested thus far, but since these channels have a fully visible final state they should allow for better sensitivity than the already performed $n p \to \tau^+\nu$ search in SK~\cite{Takhistov:2015fao}, which gave a limit of $29 \times \unit[10^{30}]{yr}$.

The invisible di-nucleon decay $n n \to\text{neutrinos}$ can be searched in a way analogous to the $\Delta B =1$ case by looking for nuclear de-excitation emission~\cite{Hagino:2018xnt}. The current best limits come from KamLAND~\cite{Araki:2005jt}, with a limit $\tau_{nn}>\unit[1.4\times 10^{30}]{yr}$. Slightly lower limits around a few~$\times \unit[10^{28}]{yr}$ have been obtained by SNO+~\cite{Anderson:2018byx} for $np\to\text{invisible}$ and $pp\to\text{invisible}$ modes. We envision that JUNO will be able to test these modes with lifetime sensitivity over an order of magnitude stronger than KamLAND, approaching $\unit[10^{31}]{yr}$. Dedicated studies of sensitivity to these modes in DUNE, JUNO, and HK would provide valuable complementary information. 

\subsection{\texorpdfstring{$\Delta B=2$}{Delta B = 2}, \texorpdfstring{$\Delta L = \pm 2$}{Delta L = +-2}}

At dimension $d=12$, we find $\Delta B=\Delta L=2$ operators, an example being $uuuuddee$. For first-generation quark-flavor indices, this simply induces the highly visible $pp\to \ell^+\ell'^+$, which has been discussed in the context of GUTs~\cite{Arnellos:1982nt}. While the electron and muon modes have been searched for in SK~\cite{Sussman:2018ylo}, the one kinematically available tau mode $pp\to e^+\tau^+$ has not. Similar to the case of $nn \to \mu^+\tau^-$ we expect that SK can probe its lifetime above $\unit[10^{31}]{yr}$ and better than the $n p \to \tau^+\nu$ search~\cite{Takhistov:2015fao}. HK will allow us to considerably improve on studies of these channels.
We note that taking the quark indices of the $d=12$ operators into account can lead to additional emission of kaons as well as other mesons in the final state, which we do not list here.

Still at dimension $d=12$, we find $\Delta B=-\Delta L=2$ operators such as $dddddd\bar e\bar e$. Here, the dominant decay mode is $nn\to \ell^-\ell^-\pi^+\pi^+$, potentially with kaons instead of pions. This competes with the loop-induced $nn\to \nu\nu$, for which experimental limits exist. Other than this invisible di-neutron decay, no two-body decay fulfils $\Delta B=-\Delta L=2$, highlighting the importance of multi-body final states and inclusive searches.

\subsection{\texorpdfstring{$\Delta B >2$}{Delta B > 2}}
\label{sec:DeltaBabove2}

Baryon number violation by more than two units has not been well explored, with sparse theoretical~\cite{Babu:2003qh} and experimental~\cite{Hazama:1994zz,Bernabei:2006tw,Albert:2017qto,Alvis:2018pne} studies.
As mentioned before, one particularly well-motivated case that falls into this category is $\Delta B = \Delta L =3$, as mediated in the SM by non-perturbative electroweak instantons. At low temperatures and energies as in nuclear decays these processes are highly suppressed, while rate calculations in high-energy collider setups are notoriously difficult~\cite{Ringwald:1989ee,Espinosa:1989qn}.
While we do not discuss the various $\Delta B >2$ final states and operators (starting at $d=15$) further, we stress that inclusive searches can easily cover these processes as well, as we describe below.

\subsection{Inclusive searches for \texorpdfstring{$\Delta B > 1$}{Delta B > 1}}

As discussed above, there is still significant untested parameter space for $\Delta B \geq 2$ processes, including about a hundred $\Delta B=2$ two-body final states with fixed kinematics (see Tabs.~\ref{tab:two-body_dinucleon_limits1},~\ref{tab:two-body_dinucleon_limits2}, and~\ref{tab:two-body_dinucleon_limits3}). While exclusive searches for these channels would clearly provide the best sensitivity, \emph{inclusive} searches again offer an interesting alternative. A particularly relevant aspect of the proposed inclusive $\Delta B =1$ searches from Sec.~\ref{sec:NtoXplusanything} to stress is that they will \emph{automatically} provide limits for $\Delta B >1$ channels. For example, an inclusive search for $p\to e^++\text{anything}$ with positron energy between $\sim\unit{MeV}$ and $m_p/2$ would also constrain $n p\to e^++\text{anything}$ or $n n p\to e^++\text{anything}$. In comparison, dedicated $\Delta B >1$ inclusive searches would allow us to widen the search window to positron energies above $m_p/2$ or look for on-shell tau leptons.

\section{Conclusions}
\label{sec:conclusions}

Baryon number violation is strongly motivated from many distinct theoretical considerations, including GUTs, supersymmetry, and baryogenesis.
Thus, probing proton decay and other baryon-number-violating processes is of fundamental importance in order to learn about physics beyond the SM.
A slew of upcoming large-scale experiments, in particular JUNO, DUNE, and Hyper-Kamiokande, will be able to explore these processes with unparalleled sensitivity.  Hence, it is crucial to make the most of these efforts (as well as existing detectors, most notably SK) and probe as much parameter space for baryon-number violation as possible.

As we highlighted in this work, nucleon decay constitutes a premier probe of baryon number violation because $\Delta B = 1$ operators will result in nucleon decay regardless of their quark or lepton flavor structure.
While most nucleon decay searches have focused on two-body decay channels (e.g.~$p \to e^+\pi^0$), as motivated by minimal GUT models and simplified EFT considerations, a vast landscape of possible baryon-number-violating processes exists beyond them. 

Nucleon decay provides a unique opportunity to test baryon-number-violating operators of very high dimensionality and even beyond $\Delta B = 1$, achieving sensitivity not available with other techniques. These operators can dominate over the conventionally considered lower-order dimension-six terms due to their lepton number and flavor quantum numbers. Going beyond the lower order, a vast amount of additional possible operators can start to contribute. Generically, such operators will result in complicated nucleon decays with a multi-body final state.

Most of the conducted nucleon decay searches have been exclusive, focusing on decay processes with a particular final state. We have identified a slew of single nucleon decay processes yet to be tested with exclusive studies. Further, exclusive searches for relatively clean channels like $p\to \ell^+\ell^-\ell^+$ or $n\to \ell^- K^+\pi^0$ could provide information indicative of $d=9$ and $10$ operators.

Exclusive studies rapidly become unfeasible to test nucleon decay beyond the simplest channels.
As we argued in this work, a particularly important venue to further explore baryon number violation in the future is through studies of \emph{inclusive} nucleon decays $N \rightarrow X +$anything (where $X$ is a light SM particle with unknown energy distribution), as well as model-independent and invisible decay searches such as $n\to$~neutrinos (which can carry away an arbitrary lepton number). Such searches allow us to constrain multiple complicated nucleon decay channels  simultaneously and broadly cover dimension $d \geq 9$ baryon-number-violating operators. More so, inclusive $\Delta B = 1$ searches will allow us to gain some insight into $\Delta B > 1$ processes as well.

For the searches outlined above, we foresee that the WC detectors SK and HK will in general provide the furthest reach in sensitivity due to their unparalleled size. However, we stress that searches involving heavier final-state charged particles, e.g.~kaons, can significantly benefit from studies in JUNO and DUNE. Further, the low-energy sensitivity of JUNO and DUNE could be particularly beneficial for nuclear de-excitation emission searches (e.g.~invisible decay channels). We are not aware of dedicated sensitivity studies related to these aspects.
Finally, let us stress that even though we envision nuclear decays as the best probe for $\Delta B \neq 0$, other baryon-number-violating searches as in meson or tau decays could provide important complementary information.

\section*{Acknowledgements}

We would like to thank Mu-Chun Chen, Huan Huang, Ed Kearns, and Henry Sobel for discussions.
The work of J.H.~is supported, in part, by the National Science Foundation under Grant No.~PHY-1620638 and PHY-1915005 and by a Feodor Lynen Research Fellowship of the Alexander von Humboldt Foundation. The work of V.T.~is supported by the U.S.~Department of Energy (DOE) Grant No.~DE-SC0009937.
This work was performed in part at the Aspen Center for Physics, which is supported by the National Science Foundation grant PHY-1607611.
J.H.~further thanks the CERN theory group for hospitality while this article was finalized.\\[2ex]

\emph{Note added:} Recently, two articles came out that bear relevance and are complimentary to our work. In Ref.~\cite{Girmohanta:2019xya} the authors estimate the branching ratios of proton and neutron into two-body and some three-body final states, induced by one of the $d=6$ operators from Eq.~\eqref{eq:dequal6}, concluding that two-body decays currently provide the best limits, in particular $p\to \ell^+ \pi^0$ and $n\to \nu \pi^0$. A similar analysis is performed in Ref.~\cite{Girmohanta:2019cjm} but for processes induced by $d=9$ operators with $\Delta B=2$ and $\Delta L =0$.
We agree with their conclusions, but stress once more that operators with $d>6$ ($d>9$) and different lepton numbers can significantly change the dominant nucleon (di-nucleon) decay modes.

\let\oldaddcontentsline\addcontentsline
\renewcommand{\addcontentsline}[3]{}
\bibliographystyle{utcaps_mod}
\bibliography{pdk}
\let\addcontentsline\oldaddcontentsline

\end{document}